\documentclass[aps,prl,reprint,superscriptaddress,nofootinbib,floatfix]{revtex4-2}

\usepackage[utf8]{inputenc}
\usepackage[T1]{fontenc}
\usepackage{amsmath,amssymb,bm}
\usepackage{graphicx}
\usepackage{xcolor}
\usepackage{hyperref}

\AtBeginDocument{%
  \setlength{\abovedisplayskip}{6pt plus 2pt minus 2pt}%
  \setlength{\belowdisplayskip}{6pt plus 2pt minus 2pt}%
  \setlength{\abovedisplayshortskip}{3pt plus 2pt}%
  \setlength{\belowdisplayshortskip}{4pt plus 2pt minus 2pt}%
}

\hypersetup{
  colorlinks=true,
  linkcolor=blue,
  citecolor=blue,
  urlcolor=blue
}

\begin{document}

\title{Microscopic Nonreciprocity-Governed Topological Boundary Flows in Active Matter}

\author{Tong Zhu}
\email{zhutong@westlake.edu.cn}
\affiliation{Center for Interdisciplinary Studies, School of Science, Westlake University, Hangzhou 310030, CHINA}
\author{Zhigang Zheng}
\email{zgzheng@hqu.edu.cn}
\affiliation{Institute of System Science, Huaqiao University, Xiamen 361021, CHINA}
\affiliation{College of Information Science and Technology, Huaqiao University, Xiamen 361021, CHINA}
\date{\today}

\begin{abstract}
We connect microscopic nonreciprocity to chiral boundary transport in a minimal frustrated Vicsek--Kuramoto model. Unlike field-level constructions, a low-angular-harmonic closure of particle dynamics yields the non-Hermitian operator. Direction-independent polar projector limits permit wave-number compactification, and the Sakaguchi phase lag reverses the polar spin weights, producing $C=\pm2$ where the closure is nonsingular and the projector remains globally isolated. Strip edge-mode propagation agrees with particle circulation. Crucially, chirality and defect tolerance are distinct: before pattern onset, boundary streams retain chirality on regular convex boundaries but are vulnerable to strong isolated defects; an interior vortex lattice supports a counterpropagating edge stream with substantially stronger defect tolerance in the tested geometries. Thus spectral topology selects a chiral boundary branch, while pattern formation supports its robust nonlinear realization.
\end{abstract}

\maketitle

\emph{Introduction.---}
Active matter organizes local propulsion into collective motion, with Vicsek-type alignment providing a route to flocking \cite{Ramaswamy2010,Marchetti2013,Vicsek1995,TonerTu1995}. Kuramoto coupling provides a complementary route through phase locking, frustrated by a Sakaguchi phase lag \cite{Kuramoto1984,Strogatz2000,Acebron2005,Sakaguchi1988}. Combining motion and phase-like orientation connects synchronization, swarming, and transport in Kuramoto--Vicsek models and related active systems \cite{Degond2013KuramotoVicsek,Chate2015Memory,LiebchenLevis2017}.

Topology connects bulk structure to boundary response, from quantum Hall systems \cite{Thouless1982,Hatsugai1993} to photonic \cite{HaldaneRaghu2008,WangPhotonic2008}, acoustic \cite{YangAcoustic2015,HeAcoustic2016}, mechanical \cite{KaneLubensky2014,SusstrunkHuber2015}, geophysical \cite{Delplace2017,Tauber2019}, non-Hermitian \cite{YaoWang2018,Kunst2018}, and active systems \cite{Ashida_Anomalous_2019,Tang_Topology_2021}. In spectral problems, Chern numbers characterize eigenvector bundles and, under suitable gap and boundary conditions, constrain edge spectral flow. Electronic Hall responses illustrate how microscopic band geometry controls transport \cite{Yao2004AHE,Qiao2010QAHGraphene}; active matter extends topological ideas to defects, curved-surface flocking, and chiral currents \cite{Shankar_Topological_2022}. Can an analogous spectral mechanism arise directly from self-organizing particle rules?

Confinement reorganizes bacterial motion into vortices and cellular order \cite{Wioland2013,Lushi2014}, while swarming colonies exhibit robust edge flows \cite{Zhang2024RobustEdge}. Such observations motivate distinguishing boundary-selected organization from a response encoded in bulk topology. Here collision boundaries provide a setting in which to test that distinction.

Our starting point is a frustrated Vicsek--Kuramoto model whose periodic-boundary instability shifts from zero to finite wave number near $\alpha=\pi/2$, separating synchronized clusters from dynamic vortex lattices \cite{Lu2025SwarmingLattice}. Above this threshold, a phase-locking ansatz associates a narrow hyperuniform regime with incompatibility between the vortex scale and the linearly selected wavelength \cite{Lu2025SwarmingLattice,LuZhu2026Hyperuniformity}.

Unlike field-level topological-fluid constructions \cite{Souslov_Topological_2019}, we obtain the spectral structure through a closure of microscopic dynamics. Rotational covariance and direction-independent polar projector limits permit wave-number compactification; phase-lag-induced reversal of the polar spin weights yields $C=\pm2$ where the closure is nonsingular and the projector remains isolated. Strip edge modes reproduce the observed circulation direction. Particle simulations further distinguish persistent chirality from defect tolerance: an interior vortex lattice supports robust edge transport that no-pattern states do not sustain under strong defects.

\begin{center}
  \centering
  \includegraphics[width=\linewidth]{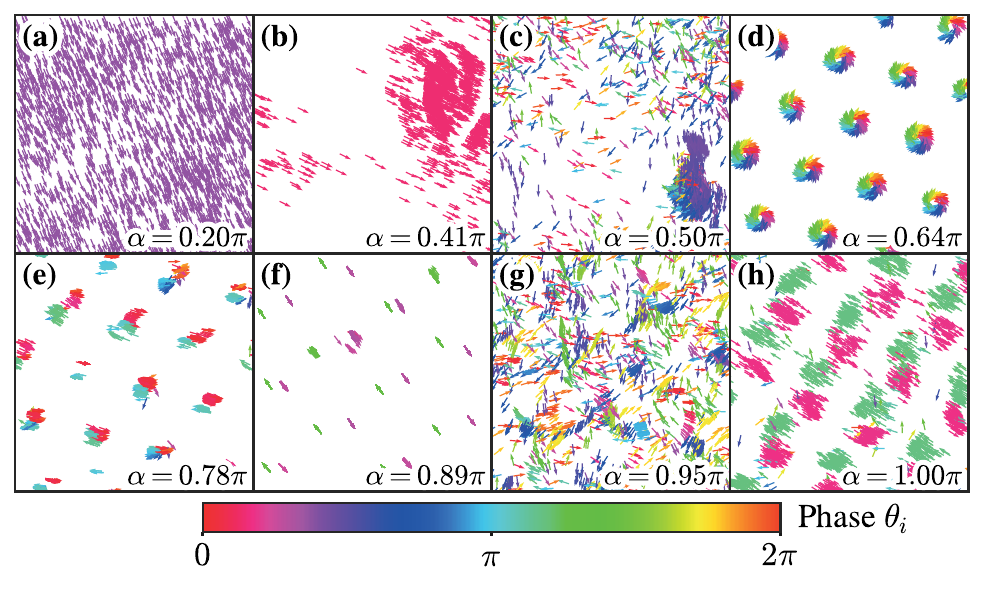}
  \refstepcounter{figure}\label{fig:prl_pbc_phase}
  \parbox{\linewidth}{\small\textbf{FIG. \thefigure.} Periodic-boundary terminal states across the phase-lag sweep for $L=7$, $N=2000$, $K=20$, $d_0=1.55$, and $v=3$, reproduced from Ref.~\cite{Lu2025SwarmingLattice}. Below the finite-wavenumber threshold the system forms synchronized clusters; above it the spectrum selects a finite pattern scale.}
\end{center}

\begin{figure*}[t]
  \centering
  \includegraphics[width=\textwidth]{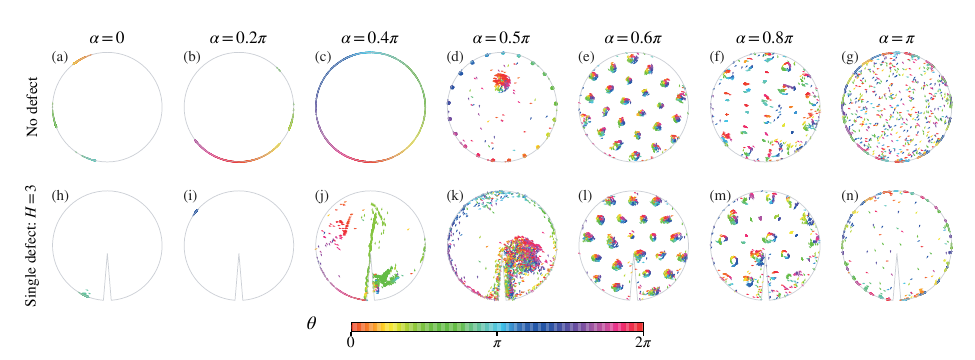}
  \caption{Circular-boundary terminal states without a defect (a--g) and with one inward triangular defect of height $H=3$ in simulation length units (h--n), for $L=7$, $N=2000$, $K=20.75$, $v=3$, and $d_0=1$; color gives $\theta$. Columns sweep $\alpha=0,0.2\pi,0.4\pi,0.5\pi,0.6\pi,0.8\pi,\pi$. The associated trajectory analysis assigns no persistent single chirality or topological protection at $\alpha=0$. For $0<\alpha<\pi/2$, an all-particle chiral stream forms without an interior pattern but is readily disrupted by the isolated defect. At $\alpha=\pi/2$, the unidirectional boundary lattice has spacing of order $d_0$ and remains defect-sensitive. For $\pi/2<\alpha<\pi$, the interior vortex lattice supports a highly coherent minority edge stream that survives the defect. At the singular phase-lag endpoint $\alpha=\pi$, trajectory data show two counterpropagating boundary-lattice streams whose spacing is associated with the limiting finite-wavenumber scale $2\pi/k_*$ rather than the threshold scale $d_0$; this bidirectional state is distinct from the chiral $\alpha=\pi/2$ lattice and carries no Chern label within the continuum closure.}
  \label{fig:prl_collision_states}
\end{figure*}
\begin{figure}[t]
  \centering
  \noindent\makebox[\linewidth][c]{%
    \begin{minipage}{0.48\linewidth}
      \centering
      \textbf{(a) $\alpha=0.4\pi$}\\[-0.2em]
      \includegraphics[width=\linewidth]{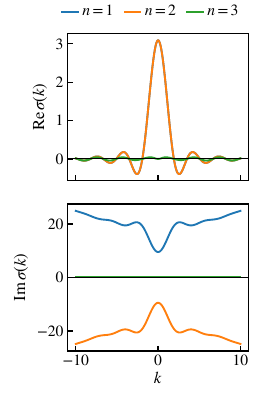}
    \end{minipage}\hspace{0.02\linewidth}%
    \begin{minipage}{0.48\linewidth}
      \centering
      \textbf{(b) $\alpha=0.6\pi$}\\[-0.2em]
      \includegraphics[width=\linewidth]{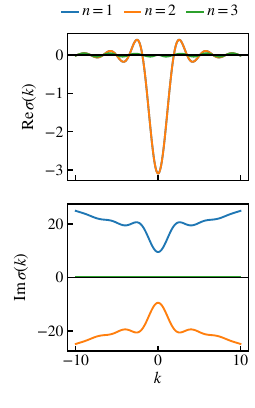}
    \end{minipage}%
  }
  \caption{Directional dispersions. Isotropy makes the $\alpha=0.4\pi$ and $0.6\pi$ cuts representative and exposes the zero- to finite-wavenumber instability transition.}
  \label{fig:prl_dispersion}
\end{figure}
\begin{figure*}[t]
  \centering
  \begin{minipage}{0.49\linewidth}
    \centering
    \textbf{(a) mixed-band Chern}\\[-0.2em]
    \includegraphics[width=\linewidth]{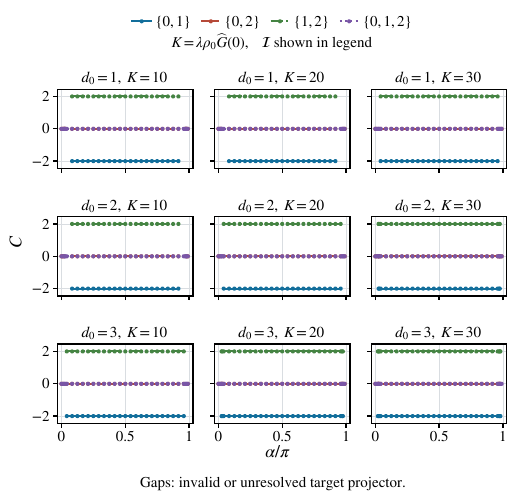}
  \end{minipage}\hfill
  \begin{minipage}{0.49\linewidth}
    \centering
    \textbf{(b) single-band Chern}\\[-0.2em]
    \includegraphics[width=\linewidth]{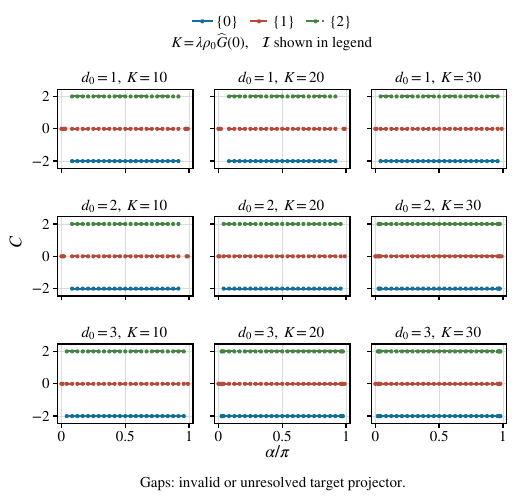}
  \end{minipage}
  \caption{Chern platforms. (a,b) Mixed- and single-band results for $v=3$, $\omega=0$, obtained from the pole formula after numerical radial spectral-isolation screening. Gaps in a curve mark an undefined or numerically unresolved target projector, including finite-wavenumber exceptional-point collisions and the singular phase-lag endpoints; they are not zero Chern values. Representative values are independently checked by the Fukui discretization.}
  \label{fig:prl_chern_platforms}
\end{figure*}
\begin{figure*}[t]
  \centering
  \includegraphics[width=0.94\textwidth]{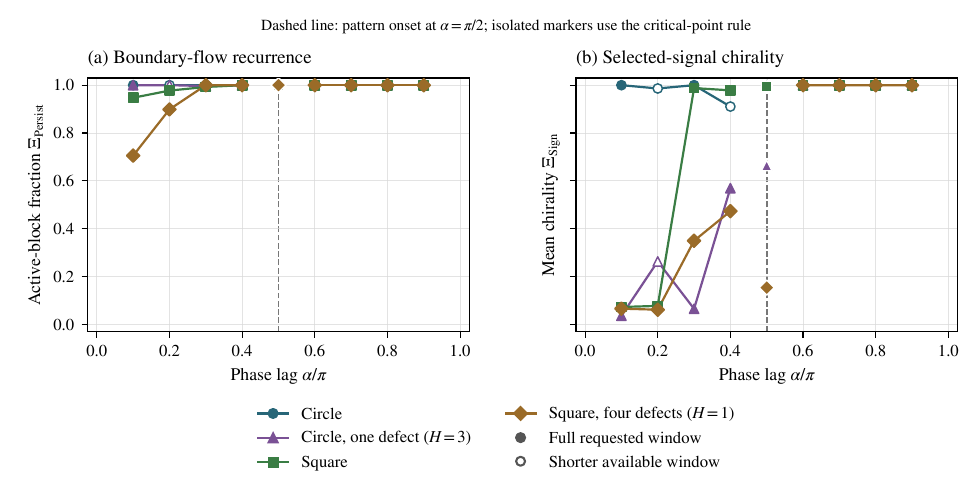}
  \caption{Single-realization, terminal-window boundary-flow diagnostics across the phase-lag sweep. (a) The recurrent active-block fraction remains high through most of the open interval. (b) Mean chirality is stable on a regular circle below pattern onset but is strongly reduced by corners or geometric defects; after the interior pattern forms, all tested geometries retain near-unit mean chirality of the selected boundary population. The isolated $\alpha=\pi/2$ markers use the critical-point selection rule and remain defect-sensitive. Filled markers reach the requested ten-circuit window, whereas open markers use shorter records. The singular phase-lag endpoints $\alpha=0,\pi$ are excluded from these single-chirality measures; definitions, coverage checks, and numerical limitations are given in the Supplemental Material \cite{Zhu2026MethodsAppendix}.}
  \label{fig:prl_boundary_stability}
\end{figure*}

\emph{Active-particle model.---}
The microscopic dynamics are
\begin{equation}
  \dot{\mathbf x}_i=v\mathbf e(\theta_i),\qquad
  \mathbf e(\theta_i)=(\cos\theta_i,\sin\theta_i),
  \label{eq:prl_position}
\end{equation}
and
\begin{equation}
  \dot{\theta}_i=\omega_i+
  \frac{K}{|\mathcal N_i|}
  \sum_{j\in\mathcal N_i}
  [\sin(\theta_j-\theta_i+\alpha)-\sin\alpha],
  \label{eq:prl_angle}
\end{equation}
for $i=1,2,\dots,N$. Here $K$ is the neighborhood-averaged coupling strength, $\omega_i=0$ is the intrinsic angular frequency, $\mathcal N_i$ contains the particles within the interaction range of particle $i$, and $\alpha$ is the frustration angle. The subtraction of $\sin\alpha$ removes the uniform angular drift of a perfectly aligned neighborhood. In the continuum equations below, the coefficient $\lambda$ denotes the effective strength after replacing the discrete average by an unnormalized kernel convolution; for a typical neighborhood population $N'$, $\lambda\simeq K/N'$ up to kernel normalization \cite{Zhu2026MethodsAppendix}.

Figures~\ref{fig:prl_pbc_phase} and \ref{fig:prl_collision_states} compare periodic and collision boundaries, while Fig.~\ref{fig:prl_boundary_stability} distinguishes persistent chirality from defect tolerance. For $0<\alpha<\pi/2$, all-particle boundary streams can retain chirality on regular convex boundaries but remain vulnerable to strong isolated defects. For $\pi/2<\alpha<\pi$, an interior vortex lattice coexists with a counterpropagating minority edge stream and supports stronger defect tolerance in the tested geometries. We call this relative bulk--edge motion ``convection'' only in a kinematic sense, not fluid-mechanical convection.

The boundary lattices at $\alpha=\pi/2$ and $\pi$ are distinct: the former is unidirectional, has spacing of order $d_0$, and remains defect-sensitive; the latter supports two counterpropagating streams whose spacing follows the limiting scale $2\pi/k_*$ from the nonsingular side, where $k_*$ maximizes the linear growth rate. At $\alpha=0$, the tested trajectories lack persistent net chirality and depend on initialization and geometry. Both phase-lag endpoints are excluded from the present Chern classification because the continuum closure is singular, not because they have zero Chern number. Trajectory diagnostics, geometry tests, and scale comparisons are detailed in the Supplemental Material \cite{Zhu2026MethodsAppendix}. We next examine the spectral mechanism underlying the observed chirality wherever the continuum projector is well defined and isolated.

\emph{Continuum spectrum.---}
The directional spectrum and Chern platforms are summarized in Figs.~\ref{fig:prl_dispersion} and \ref{fig:prl_chern_platforms}, respectively.
\raggedbottom
Let $f(\mathbf x,\theta,t)$ be the one-particle density in position $\mathbf x$ and heading $\theta$. Mean-field reduction of the $N$-particle Liouville equation gives its kinetic PDE. Motivated by the first-angular-harmonic interaction, we expand $f$ in angular modes indexed by $m$, neglect large $|m|$, and adiabatically eliminate the second harmonic as a phenomenological closure \cite{Zhu2026MethodsAppendix}. Topology belongs to this retained operator with its stated large-wave-number extension, not the full kinetic hierarchy. Below, $\mathbf k$ is the spatial wave vector.

Before linearization, define the spatial density and polarization by
\begin{equation}
  \rho(\mathbf x,t)=\int_0^{2\pi}f\,d\theta,
  \qquad
  \mathbf p(\mathbf x,t)=\int_0^{2\pi}f\,\mathbf e(\theta)\,d\theta .
  \label{eq:prl_moments}
\end{equation}
With homogeneous density $\rho_0$ and hats denoting spatial Fourier transforms, define
\begin{equation}
  \mathbf U(\mathbf k,t)=
  (\widehat{\delta\rho},\widehat{\delta p_x},\widehat{\delta p_y})^T,
  \qquad \mathbf k=(k_x,k_y),\quad k=|\mathbf k| .
  \label{eq:prl_state}
\end{equation}
For the isotropic kernel $G(r)$, define
\begin{equation}
  \widehat G(k)=\int_{\mathbb R^2}G(|\mathbf r|)e^{-i\mathbf k\cdot\mathbf r}\,d^2\mathbf r,
  \qquad G_0=\widehat G(0).
  \label{eq:prl_kernel}
\end{equation}
Retaining $\omega$ symbolically, define the closure denominator and radial coefficients
\begin{align}
  D_0&=2\omega-2\lambda\rho_0G_0\sin\alpha,\nonumber\\
  a(k)&=\frac{\lambda\rho_0}{2}\widehat G(k)\cos\alpha,\nonumber\\
  b(k)&=-\omega+\lambda\rho_0G_0\sin\alpha
  -\frac{\lambda\rho_0}{2}\widehat G(k)\sin\alpha
  +\frac{v^2k^2}{4D_0}.
  \label{eq:prl_radial_coefficients}
\end{align}
The linearized dynamics are
\begin{equation}
  \partial_t\mathbf U=M(\mathbf k)\mathbf U,\qquad
  M(\mathbf k)=
  \begin{pmatrix}
  0 & -ivk_x & -ivk_y\\
  -\frac{iv}{2}k_x & a(k) & b(k)\\
  -\frac{iv}{2}k_y & -b(k) & a(k)
  \end{pmatrix}.
  \label{eq:prl_M}
\end{equation}
For the disk kernel, $\widehat G(k)=2\pi d_0J_1(kd_0)/k$. The closure requires $D_0\ne0$; since $\omega=0$ here, $\alpha=0$ and $\pi$ are singular phase-lag endpoints rather than topologically trivial points. The generally complex eigenvalues $\sigma_n$ have growth rates $\operatorname{Re}\sigma_n$ and frequencies $\operatorname{Im}\sigma_n$. Figure~\ref{fig:prl_dispersion} shows that the dominant growth rate moves from $k=0$ at $0.4\pi$ to a finite wave number at $0.6\pi$.

\emph{Chern number from isotropy and spin.---}
Fix an isolated band or band cluster $\mathcal I$. Its right and left eigenvectors are defined biorthogonally by
\begin{equation}
  \begin{aligned}
  M|u_n^R\rangle&=\sigma_n|u_n^R\rangle,\qquad
  \langle u_n^L|M=\sigma_n\langle u_n^L|,\\
  \langle u_\ell^L|u_n^R\rangle&=\delta_{\ell n}.
  \end{aligned}
  \label{eq:prl_biorthogonal}
\end{equation}
The corresponding spectral projector is
\begin{equation}
  P_{\mathcal I}(\mathbf k)=
  \sum_{n\in\mathcal I}|u_n^R(\mathbf k)\rangle
  \langle u_n^L(\mathbf k)| .
  \label{eq:prl_projector}
\end{equation}
Writing $d$ for the exterior derivative on a compact wave-number surface $\mathcal M$, its first Chern number is
\begin{equation}
  C(P_{\mathcal I})=\frac{1}{2\pi i}\int_{\mathcal M}
  \operatorname{Tr}\!\left(P_{\mathcal I}\,dP_{\mathcal I}
  \wedge dP_{\mathcal I}\right).
  \label{eq:prl_chern}
\end{equation}
Figure~\ref{fig:prl_chern_platforms} uses the pole formula below after spectral-isolation screening, with representative values checked independently by the biorthogonal Fukui--Hatsugai--Suzuki algorithm \cite{Fukui2005}.

To identify $\mathcal M$ and evaluate $C$, parameterize the wave vector and its density--polarization rotation by
\begin{equation}
  \mathbf k=k(\cos\phi,\sin\phi),\qquad
  S(\phi)=1\oplus
  \begin{pmatrix}
    \cos\phi&-\sin\phi\\
    \sin\phi&\cos\phi
  \end{pmatrix}.
  \label{eq:prl_rotation}
\end{equation}
With $B(k)=M(k,0)$, rotational covariance implies
\begin{equation}
  \begin{aligned}
  M(k,\phi)&=S(\phi)B(k)S^{-1}(\phi),\\
  \det[M(k,\phi)-\sigma I]&=\det[B(k)-\sigma I].
  \end{aligned}
  \label{eq:prl_covariance}
\end{equation}
Thus $\sigma_n(k,\phi)=\sigma_n(k)$ along each continuously labelled branch,
with labels ambiguous at degeneracies. Define the angular generator $J$
and its density and circular-polarization eigenvectors by
\begin{equation}
  \begin{gathered}
  J=iS^{-1}\partial_\phi S,\qquad
  e_\rho=(1,0,0)^T,\qquad u_\pm=(0,1,\pm i)^T,\\
  Je_\rho=0,\qquad Ju_\pm=\pm u_\pm .
  \end{gathered}
  \label{eq:prl_spin_basis}
\end{equation}
\begin{samepage}
At large $k$, label the spin sectors by $s=0,\pm1$. Let $\sigma_s$ and
$P_s$ denote the associated eigenvalue and Riesz projector, and set
$\beta=v^2/(4D_0)$.
\begin{equation}
  \begin{aligned}
  \frac{B(k)}{k^2}&\longrightarrow i\beta J,
  &\frac{\sigma_s(k,\phi)}{k^2}&\longrightarrow i\beta s,\\
  P_s(k,\phi)&\longrightarrow P_s(\infty).
  \end{aligned}
  \label{eq:prl_infinity_data}
\end{equation}
\end{samepage}
These asymptotics are uniform in $\phi$: unscaled oscillatory eigenvalues
diverge in the imaginary direction, but each branch has a unique
direction-independent asymptotic spectral value. The polar projector limits,
also direction-independent at $k=0$, allow extension over the compactified plane
\begin{equation}
  \mathcal M=\mathbb R^2\cup\{\infty\}\simeq S^2.
  \label{eq:prl_compactification}
\end{equation}
Order eigenvalues by increasing imaginary part, $\operatorname{Im}\sigma_1\leq\operatorname{Im}\sigma_2\leq\operatorname{Im}\sigma_3$. For the first branch, let $P(k)$ be the projector of $B(k)$ and define its pole coefficient and spins
\begin{equation}
  \begin{aligned}
  b_0&=-\omega+\frac{\lambda\rho_0G_0}{2}\sin\alpha,\\
  s_0&=-\operatorname{sgn}b_0,
  \qquad s_\infty=-\operatorname{sgn}\beta .
  \end{aligned}
  \label{eq:prl_pole_data}
\end{equation}
The pole eigenvectors are $u_{s_0}$ and $u_{s_\infty}$. For $\omega=0$ and $0<\alpha<\pi$, the microscopic phase lag enforces the twist
\begin{equation}
  \begin{aligned}
  D_0b_0&=-(\lambda\rho_0G_0)^2\sin^2\alpha<0,\\
  b_0\beta&<0,\qquad s_\infty=-s_0 .
  \end{aligned}
  \label{eq:prl_twist_condition}
\end{equation}
Microscopic nonreciprocity thus reverses the spin between the compactified wave-number poles. Rotational covariance reduces the Chern integral to their spin-weight difference,
\begin{equation}
  \begin{aligned}
  C\equiv C(P_{\mathcal I})&=-\left[\operatorname{Tr}(JP(k))\right]_{0}^{\infty}
  =-(s_\infty-s_0),\\
  |C|&=2,
  \end{aligned}
  \label{eq:prl_pole_chern}
\end{equation}
up to the orientation convention, provided the target projector is isolated for every finite $k$. Opposite pole spins alone do not establish this condition: finite-$k$ exceptional points exclude portions of the open phase-lag interval. Figure~\ref{fig:prl_chern_platforms} marks invalid or unresolved samples as unavailable; at $\alpha=0,\pi$, the closure itself is singular because $D_0=0$ \cite{Zhu2026MethodsAppendix}.

\emph{Boundary spectral flow.---}
For tangential wavenumber $k_y$, Fourier blocks of $M(k_x,k_y)$ define a strip operator with bulk-like and edge-localized modes. For an edge branch, let $\gamma$ denote the growth rate and $\Omega$ the oscillation frequency. Its eigenvalue and group velocity are
\begin{equation}
  \sigma(k_y)=\gamma(k_y)+i\Omega(k_y),\qquad
  v_g=-\partial_{k_y}\Omega(k_y).
  \label{eq:prl_group_velocity}
\end{equation}
With the convention $e^{ik_y y+\sigma t}$, $e^{\gamma t}$ sets linear growth and $k_y y+\Omega t$ gives the propagation phase. In Fig.~\ref{fig:prl_flow}, $v_g<0$ on the left edge and $v_g>0$ on the right predict counterclockwise circulation around a closed collision boundary, consistent with particle simulations. This comparison assumes linear mode selection followed by nonlinear saturation: topology belongs to the spectral projector, while particle organization supports the finite-amplitude flow.

\noindent\begin{minipage}{\linewidth}
  \centering
  \includegraphics[width=\linewidth]{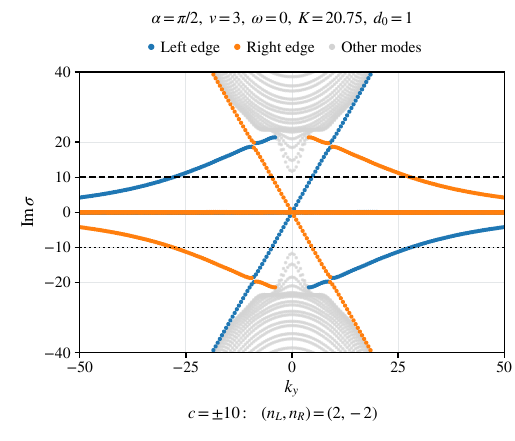}
  \refstepcounter{figure}\label{fig:prl_flow}
  \parbox{\linewidth}{\small\textbf{FIG. \thefigure.} Finite-strip spectrum at $\alpha=\frac{\pi}{2}$, with $v=3$, $\omega=0$, $K=20.75$, and $d_0=1$ matched to the circular-particle runs. The empirical levels $\operatorname{Im}\sigma=\pm10$ lie in bulk line gaps at this parameter value. The multi-$\alpha$ comparison and finite-grid diagnostics are given in the Supplemental Material.}
\end{minipage}
\emph{Discussion.---}
Microscopic nonreciprocity produces $C=\pm2$ through a polar spin twist in the nonsingular, spectrally isolated regimes of the continuum closure. Pattern formation does not create this Chern phase; it supports its robust nonlinear realization. Before pattern onset, persistent chirality on convex boundaries remains vulnerable to strong defects. An interior vortex lattice replenishes scattered edge carriers and supports near-unit mean chirality of the selected boundary population in the tested geometries. The singular phase-lag endpoints remain outside this classification. Distinguishing spectral mode selection from pattern-supported defect tolerance suggests a broader bulk--boundary principle for self-organized transport in active matter.

\begin{acknowledgments}
We thank Profs. Hong Qian and Leihan Tang for guidance and valuable suggestions, and Yichen Lu for helpful discussions and assistance. This work was supported in part by the National Natural Science Foundation of China under Grants No. 12375031 and No. 11875135.
\end{acknowledgments}

\setlength{\bibsep}{0pt}
\bibliography{bibliography}

\end{document}

\maketitle

\section{Purpose and Scope}

This appendix gives a self-contained derivation of the continuum and
linear-spectral framework used for the frustrated Vicsek--Kuramoto
particle model. The aim is to connect the microscopic particle dynamics
to: (i) a kinetic equation for the one-particle distribution, (ii) a
closed density--polarization hydrodynamic model, (iii) the non-Hermitian
linear dispersion matrix that selects pattern-forming wavelengths, and
(iv) the Chern-number and strip-spectral-flow algorithms used to diagnose
topological boundary transport.

The derivation adopts a low-angular-harmonic closure ansatz for a linear
and weakly nonlinear framework. The Chern number and strip spectral flow are statements about
the linearized operator and its spectral projectors. The nonlinear
particle simulations are then interpreted through a saturation ansatz:
linear growth selects the relevant wavelength, branch, and topology, while
nonlinear terms arrest the growth and organize the finite-amplitude
particle state.

The closed density--polarization theory derived below requires the
adiabatically eliminated second harmonic to have a nonzero algebraic
denominator. In the homogeneous state this condition is $D_0\ne0$. Since
the main text sets $\omega=0$, the phase-lag endpoints $\alpha=0$ and $\alpha=\pi$
have $D_0=0$ and are excluded from the closed dispersion matrix and all
subsequent spectral-projector and Chern-number constructions. The
microscopic particle dynamics remain meaningful at these phase-lag endpoints, but
the present continuum closure does not assign them a spectrum or a Chern
number.

\section{Kinetic Description}

\subsection{One-Particle Distribution}

Let $f(\mathbf x,\theta,t)$ be the one-particle distribution in position
and heading angle. Define
\begin{equation}
  \mathbf u(\mathbf x,\theta,t)=v(\cos\theta,\sin\theta),
  \qquad
  g(\mathbf x,\theta,t)=\dot\theta .
\end{equation}
The kinetic continuity equation is
\begin{equation}
  \partial_t f+\nabla_{\mathbf x}\cdot(f\mathbf u)+\partial_\theta(fg)=0.
  \label{eq:kinetic_continuity}
\end{equation}
We expand the angular dependence as
\begin{equation}
  f(\mathbf x,\theta,t)=\sum_{m=-\infty}^{\infty}f_m(\mathbf x,t)e^{im\theta},
  \qquad
  f_m=\frac{1}{2\pi}\int_0^{2\pi}f e^{-im\theta}\,d\theta .
  \label{eq:angular_modes}
\end{equation}
Here $m\in\mathbb Z$ is the angular-harmonic index. In the spatial
Fourier analysis below, bold $\mathbf k$ denotes the wave vector and
$k=|\mathbf k|$ its magnitude.

For a nonlocal interaction kernel $G$, define
\begin{equation}
  (\mathcal G h)(\mathbf x)=
  \int d^2\mathbf x'\,G(|\mathbf x'-\mathbf x|)\,h(\mathbf x'),
  \qquad
  F_m:=\mathcal G f_m .
\end{equation}

\subsection{Relation Between Particle and Continuum Couplings}

The particle simulations use the neighborhood-averaged coupling
\begin{equation}
  \dot\theta_i=\omega_i+
  \frac{K}{|\mathcal N_i|}
  \sum_{j\in\mathcal N_i}A_{ij}
  [\sin(\theta_j-\theta_i+\alpha)-\sin\alpha],
  \label{eq:particle_coupling_appendix}
\end{equation}
where $K$ is the numerical coupling strength. If the interaction disk has
radius $d_0$, the typical number of particles sampled by the disk is
\begin{equation}
  N'\simeq \rho_0\pi d_0^2
\end{equation}
for a homogeneous state with particle density $\rho_0$.

For the binary metric neighborhood used here, $A_{ij}=1$ for
$j\in\mathcal N_i$. We use the unnormalized disk kernel
$G(r)=\mathbf 1_{r\le d_0}$, so that
$\widehat G(0)=\pi d_0^2$. In the homogeneous-reference mean-field
setting, replacing the discrete neighborhood average by a continuous
average means
\begin{align}
 &\frac{1}{|\mathcal N_i|}\sum_{j\in\mathcal N_i}
 [\sin(\theta_j-\theta_i+\alpha)-\sin\alpha]
 \nonumber\\
 &\quad\longrightarrow
 \frac{1}{\widehat G(0)}
 \int d^2\mathbf x'\,G(|\mathbf x'-\mathbf x|)
 \int_0^{2\pi}
 [\sin(\theta'-\theta+\alpha)-\sin\alpha]
 \frac{f(\mathbf x',\theta',t)}{\rho_0}\,d\theta'.
 \label{eq:discrete_to_continuous_neighborhood_average}
\end{align}
Thus $1/\widehat G(0)$ supplies the spatial area average, while
$f/\rho_0$ accounts for the number-density convention used for $f$.
Absorbing these normalization factors into the coefficient of the
unnormalized convolution gives
\begin{equation}
 \boxed{\lambda=\frac{K}{\rho_0\widehat G(0)}
 =\frac{K}{\rho_0\pi d_0^2}\simeq\frac{K}{N'}},
 \qquad \lambda\rho_0\widehat G(0)=K .
 \label{eq:particle_continuum_coupling_normalization}
\end{equation}
This makes explicit the $K$--$\lambda$ convention of the main text:
$K$ remains the neighborhood-averaged particle coupling, whereas
$\lambda$ multiplies the unnormalized number-density convolution.
In particular, at $\omega=0$ the homogeneous closure denominator below
is $D_0=-2K\sin\alpha$.

This convention freezes the neighborhood population at its homogeneous
reference value $N'\simeq\rho_0\widehat G(0)$. Retaining the instantaneous
local population instead would give the coefficient
$K/(\mathcal G\rho)(\mathbf x,t)$ wherever that population is nonzero.
The two prescriptions agree upon linearization about
$\rho=\rho_0$, $\mathbf p=0$ in a translation-invariant bulk, but need
not agree for finite-amplitude density variations or a neighborhood
truncated by a boundary. The closed equations below use the stated
homogeneous-reference convention.

\subsection{Mode Equations}

The kinetic equation contains self-propulsion, intrinsic rotation, and
alignment contributions. The propulsion and rotation parts give
\begin{align}
  \left.\partial_t f_m\right|_{\rm pro}
  &=
  -\frac{v}{2}\left[
  \partial_{x_1}(f_{m+1}+f_{m-1})
  +i\partial_{x_2}(f_{m+1}-f_{m-1})
  \right],\\
  \left.\partial_t f_m\right|_{\rm rot}
  &=-im\omega f_m .
\end{align}
The alignment drift is written as
\begin{equation}
  g(\mathbf x,\theta,t)=\omega+g_{\rm ali}(\mathbf x,\theta,t),
\end{equation}
with
\begin{equation}
  g_{\rm ali}=
  \lambda\int d^2\mathbf x'\,G(|\mathbf x'-\mathbf x|)
  \int_0^{2\pi}
  [\sin(\theta'-\theta+\alpha)-\sin\alpha]
  f(\mathbf x',\theta',t)\,d\theta' .
\end{equation}
Projecting onto angular harmonics gives
\begin{equation}
  \left.\partial_t f_m\right|_{\rm ali}
  =
  -m\lambda\pi\left[
  e^{i\alpha}F_{-1}f_{m+1}
  -e^{-i\alpha}F_1f_{m-1}
  \right]
  +2\pi i m\lambda\sin\alpha\,F_0f_m .
\end{equation}
Thus
\begin{align}
  \partial_t f_m
  &=
  -\frac{v}{2}\left[
  \partial_{x_1}(f_{m+1}+f_{m-1})
  +i\partial_{x_2}(f_{m+1}-f_{m-1})
  \right]
  -im\omega f_m \nonumber\\
  &\quad
  -m\lambda\pi\left[
  e^{i\alpha}F_{-1}f_{m+1}
  -e^{-i\alpha}F_1f_{m-1}
  \right]
  +2\pi i m\lambda\sin\alpha\,F_0f_m .
  \label{eq:mode_equation}
\end{align}

\section{Hydrodynamic Closure}

\subsection{Macroscopic Fields}

The density and polarization are
\begin{equation}
  \rho(\mathbf x,t)=\int_0^{2\pi}f\,d\theta=2\pi f_0,
\end{equation}
\begin{equation}
  \mathbf p=(p_x,p_y)=
  \int_0^{2\pi}f(\cos\theta,\sin\theta)\,d\theta
  =
  \pi(f_1+f_{-1},\,i(f_1-f_{-1})).
\end{equation}
The second angular harmonic is written as
\begin{equation}
  Q=2\pi f_{-2}=Q_x+iQ_y .
\end{equation}
It represents the leading nematic correction beyond the density and
polarization fields.

\subsection{Low-Order Closure}

We adopt the low-order closure ansatz of the main text: angular modes with
$|m|\ge3$ are neglected and $\partial_t f_{\pm2}\approx0$, so $Q_x,Q_y$
are eliminated algebraically in terms of $\rho,\mathbf p$.
The condition $D_0\ne0$ ensures algebraic invertibility, not fast damping:
the homogeneous $k=0$ second-harmonic eigenvalues are $\mp iD_0$.
The ensuing topology belongs to this retained three-field operator; its
extension to arbitrarily large $k$ is an effective-theory UV assumption,
not a convergence claim for the full angular-harmonic hierarchy.

The density equation is
\begin{equation}
  \partial_t\rho+v(\partial_{x_1}p_x+\partial_{x_2}p_y)=0.
  \label{eq:density_equation}
\end{equation}
The polarization equations are
\begin{align}
  \partial_t p_x
  &=
  -\frac{v}{2}\partial_{x_1}\rho
  -\frac{v}{2}(\partial_{x_1}Q_x+\partial_{x_2}Q_y)
  -\omega p_y \nonumber\\
  &\quad
  +\frac{\lambda}{2}\rho[
  (\mathcal Gp_x)\cos\alpha-(\mathcal Gp_y)\sin\alpha]
  +\lambda\sin\alpha(\mathcal G\rho)p_y \nonumber\\
  &\quad
  -\frac{\lambda}{2}
  \left[
  (Q_x\mathcal Gp_x+Q_y\mathcal Gp_y)\cos\alpha
  +(-Q_x\mathcal Gp_y+Q_y\mathcal Gp_x)\sin\alpha
  \right],
  \label{eq:px_equation}
\end{align}
\begin{align}
  \partial_t p_y
  &=
  -\frac{v}{2}\partial_{x_2}\rho
  -\frac{v}{2}(\partial_{x_1}Q_y-\partial_{x_2}Q_x)
  +\omega p_x \nonumber\\
  &\quad
  +\frac{\lambda}{2}\rho[
  (\mathcal Gp_x)\sin\alpha+(\mathcal Gp_y)\cos\alpha]
  -\lambda\sin\alpha(\mathcal G\rho)p_x \nonumber\\
  &\quad
  -\frac{\lambda}{2}
  \left[
  (-Q_x\mathcal Gp_y+Q_y\mathcal Gp_x)\cos\alpha
  -(Q_x\mathcal Gp_x+Q_y\mathcal Gp_y)\sin\alpha
  \right].
  \label{eq:py_equation}
\end{align}

Define
\begin{equation}
  D(\mathbf x,t)=
  2\omega-2\lambda\sin\alpha\,(\mathcal G\rho)(\mathbf x,t).
\end{equation}
The quasistationary solution for the second harmonic is
\begin{align}
  Q_x&=
  \frac{1}{D}
  \left\{
  \frac{v}{2}(\partial_{x_1}p_y+\partial_{x_2}p_x)
  -\lambda[(\mathcal Gp_x)p_x-(\mathcal Gp_y)p_y]\sin\alpha \right. \nonumber\\
  &\hspace{5.8em}\left.
  -\lambda[(\mathcal Gp_x)p_y+(\mathcal Gp_y)p_x]\cos\alpha
  \right\},
  \label{eq:Qx}
\end{align}
\begin{align}
  Q_y&=
  -\frac{1}{D}
  \left\{
  \frac{v}{2}(\partial_{x_1}p_x-\partial_{x_2}p_y)
  -\lambda[(\mathcal Gp_x)p_x-(\mathcal Gp_y)p_y]\cos\alpha \right. \nonumber\\
  &\hspace{5.8em}\left.
  +\lambda[(\mathcal Gp_x)p_y+(\mathcal Gp_y)p_x]\sin\alpha
  \right\}.
  \label{eq:Qy}
\end{align}
Substitution of \eqref{eq:Qx}--\eqref{eq:Qy} into
\eqref{eq:density_equation}--\eqref{eq:py_equation} gives a closed
density--polarization model. This construction requires
$D(\mathbf x,t)\ne0$. If $D=0$, the adiabatic equations for $Q_x,Q_y$
cannot be inverted, and this three-field closure is not defined.

\section{Linear Stability Analysis}

\subsection{Homogeneous State}

Linearize around
\begin{equation}
  \rho=\rho_0,\qquad \mathbf p=0,
\end{equation}
with perturbations
\begin{equation}
  \rho=\rho_0+\delta\rho,\qquad
  p_x=\delta p_x,\qquad
  p_y=\delta p_y .
\end{equation}
The convolution diagonalizes in Fourier space:
\begin{equation}
  \widehat{\mathcal Gh}(\mathbf k)=\widehat G(k)\widehat h(\mathbf k),
\end{equation}
\begin{equation}
  \widehat G(k)=2\pi\int_0^\infty rG(r)J_0(kr)\,dr,
  \qquad
  \widehat G(0)=\int_{\mathbb R^2}G(|\mathbf r|)\,d^2\mathbf r .
\end{equation}
Define
\begin{equation}
  D_0=2\omega-2\lambda\sin\alpha\,\rho_0\widehat G(0).
\end{equation}
All formulas obtained by eliminating the second harmonic require
$D_0\ne0$. For the parameters of the main text, $\omega=0$, so
\begin{equation}
  D_0=-2\lambda\rho_0\widehat G(0)\sin\alpha
\end{equation}
vanishes at $\alpha=0$ and $\alpha=\pi$. At these two singular phase-lag endpoints,
the linear closure below contains divergent $D_0^{-1}$ coefficients and
does not define a finite dispersion matrix.
To linear order,
\begin{equation}
  Q_x\approx\frac{v}{2D_0}(\partial_{x_1}p_y+\partial_{x_2}p_x),
  \qquad
  Q_y\approx-\frac{v}{2D_0}(\partial_{x_1}p_x-\partial_{x_2}p_y).
\end{equation}

\subsection{Dispersion Matrix}

Let
\begin{equation}
  \mathbf U(\mathbf k,t)=
  (\widehat{\delta\rho},\widehat{\delta p_x},\widehat{\delta p_y})^T .
\end{equation}
The linearized equation is
\begin{equation}
  \partial_t\mathbf U=M(\mathbf k)\mathbf U,
\end{equation}
with
\begin{equation}
\resizebox{\linewidth}{!}{$
M(\mathbf k)=
\begin{pmatrix}
0 & -ivk_x & -ivk_y\\[6pt]
-\frac{iv}{2}k_x & \frac{\lambda\rho_0}{2}\widehat G(k)\cos\alpha &
-\omega+\lambda\rho_0\widehat G(0)\sin\alpha-\frac{\lambda\rho_0}{2}\widehat G(k)\sin\alpha+\frac{v^2}{4D_0}k^2\\[8pt]
-\frac{iv}{2}k_y &
\omega-\lambda\rho_0\widehat G(0)\sin\alpha+\frac{\lambda\rho_0}{2}\widehat G(k)\sin\alpha-\frac{v^2}{4D_0}k^2 &
\frac{\lambda\rho_0}{2}\widehat G(k)\cos\alpha
\end{pmatrix}.
$}
\label{eq:full_dispersion_matrix}
\end{equation}
The growth rates $\sigma(\mathbf k)$ are determined by
\begin{equation}
  \det(M(\mathbf k)-\sigma I)=0.
\end{equation}
If $\max_n \operatorname{Re}\sigma_n(k)>0$ over a finite interval of
wavenumbers, the homogeneous state is linearly unstable and the selected
unstable mode initiates pattern formation.

Equations \eqref{eq:full_dispersion_matrix} and
\eqref{eq:compact_matrix}, and hence their eigenvalues, are understood
only for $D_0\ne0$. In particular, for $\omega=0$ no spectral statement is
made at $\alpha=0$ or $\alpha=\pi$ within this closure.

It is convenient to write
\begin{equation}
M(\mathbf k)=
\begin{pmatrix}
0 & -ivk_x & -ivk_y\\[4pt]
-\dfrac{iv}{2}k_x & a(k) & b(k)\\[4pt]
-\dfrac{iv}{2}k_y & -b(k) & a(k)
\end{pmatrix},
\qquad k=\sqrt{k_x^2+k_y^2},
\label{eq:compact_matrix}
\end{equation}
where
\begin{equation}
  a(k)=\frac{\lambda\rho_0}{2}\widehat G(k)\cos\alpha,
\end{equation}
\begin{equation}
  b(k)=
  -\omega+\lambda\rho_0\widehat G(0)\sin\alpha
  -\frac{\lambda\rho_0}{2}\widehat G(k)\sin\alpha
  +\frac{v^2}{4D_0}k^2 .
\end{equation}
For a disk-like kernel,
\begin{equation}
  \widehat G(k)=2\pi d_0\frac{J_1(kd_0)}{k}.
\end{equation}

\section{Phase-Locking Ansatz and the Nonlinear Vortex Scale}

The hydrodynamic matrix determines the linearly selected wavelength, but
the final particle pattern is a finite-amplitude nonlinear state. The
periodic-boundary phenomenology therefore uses a separate
phase-locking ansatz to estimate the vortex scale selected by local
particle motion after saturation.

The ansatz is phenomenological. In a saturated local vortex, neighboring
particles are assumed to be phase locked: their heading phases need not
be equal, but their phase velocities are approximately equal. Thus, for a
local cluster,
\begin{equation}
  \dot\theta_i\simeq\Omega_{\rm pl}
  \qquad
  \text{for particles in the same locked vortex.}
\end{equation}
If the locked cluster samples the headings around a full local orbit, the
microscopic coupling gives
\begin{align}
  \Omega_{\rm pl}
  &\simeq
  \frac{K}{2\pi}\int_0^{2\pi}
  [\sin(\theta'-\theta+\alpha)-\sin\alpha]\,d\theta'
  \nonumber\\
  &=-K\sin\alpha .
  \label{eq:phase_locking_frequency_appendix}
\end{align}
The self-propulsion speed $v$ and the locked angular speed determine a
circulation radius
\begin{equation}
  r_{\rm pl}\simeq \frac{v}{|\Omega_{\rm pl}|},
\end{equation}
and hence an effective phase-locking vortex diameter
\begin{equation}
  \ell_{\rm pl}=2r_{\rm pl}
  \simeq
  \frac{2v}{K|\sin\alpha|}.
  \label{eq:phase_locking_length_appendix}
\end{equation}
For the main parameter range $0<\alpha<\pi$, this reduces to
$\ell_{\rm pl}\simeq2v/(K\sin\alpha)$.

This length is not the same object as the linear instability wavelength
\begin{equation}
  \ell_{\rm lin}=\frac{2\pi}{k_\ast},
\end{equation}
where $k_\ast$ maximizes the linear growth rate
$\operatorname{Re}\sigma(k)$. The phase-locking ansatz estimates how a
nonlinear vortex cluster wants to rotate once it has saturated, whereas
linear stability theory determines which wavelength first becomes
unstable. Ordinary swarming lattices occur when these two scales can be
accommodated simultaneously. Orientation-modulated hyperuniformity is
interpreted as the narrow post-onset regime in which
$\ell_{\rm pl}$ and $\ell_{\rm lin}$ are incompatible, so phase-locked
local rotation persists but cannot tile the domain into a regular
hexagonal vortex lattice.

The topological analysis in this appendix is independent of the detailed
nonlinear locking mechanism: it uses the homogeneous-state linear
projector and its line gap. The phase-locking ansatz only explains how
the saturated particle state accommodates, or fails to accommodate, the
length scale selected by the linear spectrum.

\section{Measuring Long-Time Boundary-Flow Stability}
\label{sec:particle_state_robustness}

This section measures the long-time stability of the particle flow near the
boundary.  Exactly two quantities are calculated.  First,
$\Xi_{\rm Persist}$ is the fraction of time blocks in which a valid directional
boundary signal is present.  Second, $\Xi_{\rm Sign}$ measures whether the mean
chirality of that signal keeps the same sign over the full observation window.
The calculation allows relay transport: the particle carrying the signal may
change from one saved frame to the next.  Every operation used to construct
these two quantities is stated below.

For the data analyzed here, the particle trajectories show the onset of the
interior pattern at $\alpha=\pi/2$.  We therefore use one particle-selection
rule for $0<\alpha<\pi/2$, another for $\pi/2<\alpha<\pi$, and report the
exact $\alpha=\pi/2$ state separately.

\subsection{Basic Saved Quantities and Boundary Coordinates}

We use two indices throughout this section.  The subscript
$i=1,\ldots,N$ labels a particle, where $N$ is the number of particles.  The
superscript $n=0,\ldots,F-1$ labels a saved frame, where $F$ is the number of
saved frames in the HDF5 file.  Thus a superscript $n$ is a time label, not a
power.  For example, $\mathbf X_i^n$ is the position of particle $i$ in saved
frame $n$.  Additional roman subscripts, such as ``$b$'' for boundary and
``$\mathrm{free}$'' for free flight, are descriptive labels rather than running
indices.

At saved frame $n$, the HDF5 file provides the position and heading angle of
particle $i$,
\begin{equation}
  \mathbf X_i^n=(x_i^n,y_i^n),
  \qquad
  \theta_i^n,
  \qquad
  \Delta t_s=\Delta t\,n_{\rm snap},
  \label{eq:boundary_raw_data}
\end{equation}
where $\Delta t_s$ is the time between two saved frames.  The simulation uses
the same variable $\theta_i^n$ as the microscopic phase and as the particle
heading.  The symbol $d_0$ below is the particle interaction distance; it is
different from the continuum denominator $D_0$.  The free-flight heading and
velocity are
\begin{equation}
  \mathbf u_i^n=(\cos\theta_i^n,\sin\theta_i^n),
  \qquad
  \mathbf v_{i,\mathrm{free}}^n=v\mathbf u_i^n.
  \label{eq:boundary_heading_velocity}
\end{equation}
If a file contains $F$ saved frames and $N$ particles, the two stored tables
are reshaped as
\begin{equation}
 \mathtt{positionX}\in\mathbb R^{FN\times2}
 \longrightarrow \{\mathbf X_i^n\}_{0\le n<F,\,1\le i\le N},
 \qquad
 \mathtt{phaseTheta}\in\mathbb R^{FN\times1}
 \longrightarrow \{\theta_i^n\}_{0\le n<F,\,1\le i\le N}.
 \label{eq:hdf5_boundary_reshape}
\end{equation}
Rows are not reordered, so row $i$ in every frame refers to the same particle.
The code checks that the position table has two columns, the heading table has
one column, both tables have the same number of rows, and that this number is
divisible by $N$.  The HDF5 schema does not store an iteration number.  The
analysis therefore uses the actual number of saved frames $F$, the saved-frame
spacing $\Delta t_s$, and the last saved frame in each file.

We orient the boundary counterclockwise.  For every particle position,
$\Pi$ returns quantities evaluated at the closest valid point on the boundary:
\begin{equation}
  \Pi(\mathbf X_i^n)
  =\bigl(d_i^n,s_i^n,\mathbf t_i^n,\kappa_{b,i}^n,e_i^n\bigr).
  \label{eq:boundary_projection_tuple}
\end{equation}
Every quantity on the right-hand side belongs to particle $i$ in saved frame
$n$.  Specifically, $d_i^n$ is the shortest distance from $\mathbf X_i^n$ to
the boundary, $s_i^n\in[0,P)$ is the counterclockwise arclength of the closest
boundary point, and $\mathbf t_i^n$ is the counterclockwise unit tangent at that
point.  In $\kappa_{b,i}^n$, the roman subscript $b$ means ``boundary,'' while
$i$ and $n$ retain their particle and saved-frame meanings; this quantity is
the signed curvature at the projected boundary point.  Finally, $e_i^n$ is the
label of the smooth boundary segment containing that point, and $P$ is the full
perimeter.  A projection is rejected if two boundary points are equally close
within $10^{-7}d_0$.  For a polygon or an arc--line boundary, we also remove a
length $0.05d_0$ from each side of every corner or join.  This prevents an
ambiguous tangent from being assigned at a nonsmooth point.  For a circle,
only the center has a nonunique closest point.  The signed tangential heading
component is
\begin{equation}
  q_i^n=\mathbf u_i^n\cdot\mathbf t_i^n.
  \label{eq:boundary_q_definition}
\end{equation}
Thus $q_i^n>0$ denotes counterclockwise heading, $q_i^n<0$ denotes clockwise
heading, and $vq_i^n$ is the tangential component of the free-flight velocity.
Because a collision can change the actual displacement, the sign of $q_i^n$
is checked against the measured change in $s_i^n$ below.

After removing the corner and join intervals, the retained boundary length is
$P_{\rm reg}=\sum_eL_e$.  Each retained smooth segment of length $L_e$ is
divided into
$N_e=\lceil2L_e/d_0\rceil$ boundary cells $I_g$.  Cell $g$ has length
$\ell_g=L_e/N_e\le d_0/2$, and
\begin{equation}
  \sum_g\ell_g=P_{\rm reg}.
  \label{eq:boundary_gate_lengths}
\end{equation}
No cell crosses a corner or join.  Cells are stored in their physical order
around the closed boundary.  When we test the total connected active length,
we perform the calculation twice: once with the physical periodic ordering and
once with adjacency broken at every removed corner or join interval.

\subsection{Selecting Particles Near the Boundary}

We first remove headings whose tangential sign cannot be resolved reliably:
\begin{equation}
  |q_i^n|\ge q_0,
  \qquad q_0=0.05.
  \label{eq:q0_dead_band}
\end{equation}
The value $q_0=0.05$ is only a sign-resolution threshold.  It does not require
the heading to be mainly tangential, and $|q_i^n|$ is not used as a weight.

For $0<\alpha<\pi/2$, the observed boundary flow is broad in the direction
normal to the wall.  We therefore apply no fixed cutoff to $d_i^n$.  All valid
projections satisfying Eq.~\eqref{eq:q0_dead_band} are candidates; the nearest
candidate in each boundary cell will be selected below.  The exact
$\alpha=\pi/2$ state uses this same rule and is plotted as a separate point.

For $\pi/2<\alpha<\pi$, interior vortices can pass close to the boundary.  We
use the heading dynamics to remove particles whose curved motion can be
explained without the wall.  At frame $n$, the neighbors of particle $i$ are
\begin{equation}
  \mathcal N_i^n=
  \left\{j\ne i:
  \left\|\mathbf X_j^n-\mathbf X_i^n\right\|\le d_0\right\}.
  \label{eq:boundary_neighbor_set}
\end{equation}
Using these neighbors, the heading angular velocity is recalculated from the
same discrete equation used in the simulation:
\begin{equation}
  \Omega_i^n=
  \begin{cases}
  \displaystyle
  \omega_i+K\left[
  \frac{1}{|\mathcal N_i^n|}
  \sum_{j\in\mathcal N_i^n}
  \sin(\theta_j^n-\theta_i^n+\alpha)-\sin\alpha
  \right],
  & |\mathcal N_i^n|>0,\\[1.1em]
  \omega_i,& |\mathcal N_i^n|=0.
  \end{cases}
  \label{eq:boundary_free_angular_velocity}
\end{equation}
The code evaluates Eq.~\eqref{eq:boundary_free_angular_velocity} directly from
the saved state.  It does not estimate $\Omega_i^n$ by differencing
$\theta_i^n$, because a collision can produce a jump between saved frames.

Define
\begin{equation}
  \kappa_{d,i}^n=
  \frac{\kappa_{b,i}^n}
       {1-\kappa_{b,i}^n d_i^n}
  \label{eq:parallel_curve_curvature}
\end{equation}
as the curvature of the curve at distance $d_i^n$ from the wall.  For motion
that is approximately tangent to such a constant-distance curve, the inward
acceleration required to follow the wall is
$\kappa_{d,i}^n v^2(q_i^n)^2$, while free heading rotation supplies
$v\Omega_i^n q_i^n$.  We define their difference divided by $v$ as
\begin{equation}
  W_i^n=
  \kappa_{d,i}^n v(q_i^n)^2-\Omega_i^n q_i^n.
  \label{eq:wall_requirement_filter}
\end{equation}
For $\pi/2<\alpha<\pi$, a particle is retained only when
\begin{equation}
  d_i^n\le d_0,
  \qquad
  1-\kappa_{b,i}^n d_i^n>0,
  \qquad
  W_i^n>\varepsilon_W,
  \qquad
  \varepsilon_W=0.02\frac{v}{d_0}.
  \label{eq:high_alpha_candidate_filter}
\end{equation}
Here $\varepsilon_W=0.02v/d_0$.  On a straight segment $\kappa_b=0$; on a
circular arc of radius $R$, $\kappa_b=1/R$.  The acceleration comparison is
valid only for nearly tangential motion at nearly fixed wall distance.  Thus
$W_i^n$ is used only to select particles; it is not interpreted as a measured
force.  The exact $\alpha=\pi/2$ calculation does not apply this $W$ filter.

\begin{figure}[H]
\centering
\includegraphics[width=\textwidth]{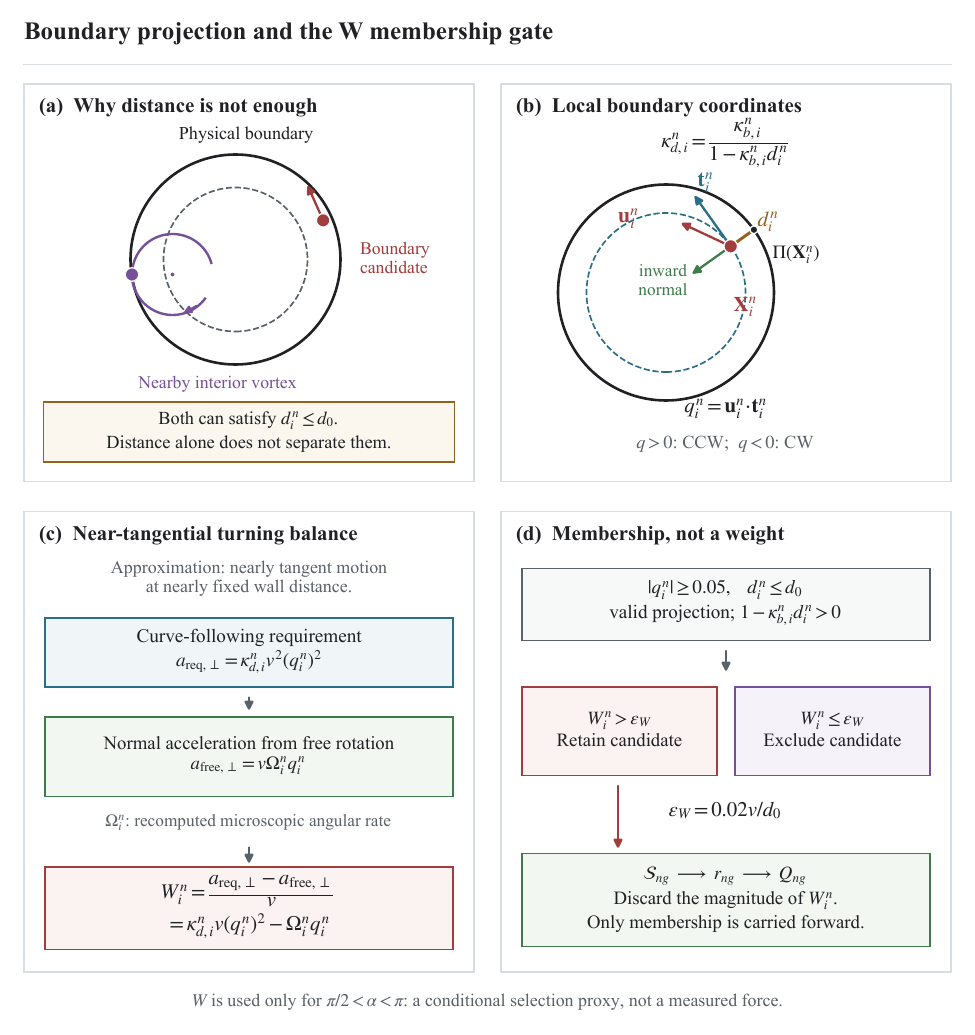}
\caption{Code-drawn construction of the boundary-selection quantity $W_i^n$.
(a) A distance cutoff alone cannot distinguish wall-supported motion from an
interior vortex that passes near the wall.  (b) Closest-point projection gives
the local tangent, normal, signed tangential component $q_i^n$, and offset
curvature $\kappa_{d,i}^n$.  (c) The curvature-following requirement is
compared with the normal acceleration already supplied by the free heading
dynamics; their difference per unit speed is $W_i^n$.  (d) $W_i^n$ is used
only as a membership gate.  Its magnitude is discarded before the cell,
relay, and long-time statistics are calculated.  The drawing states the
geometric assumptions used by the filter; it is not a simulation snapshot.}
\label{fig:boundary_projection_w_schematic}
\end{figure}

Figure~\ref{fig:boundary_projection_w_schematic} shows the quantitative origin
of Eqs.~\eqref{eq:parallel_curve_curvature}--\eqref{eq:wall_requirement_filter}.
It also makes explicit why the required curvature term contains $(q_i^n)^2$
but the free-turning term contains the signed factor $q_i^n$.

Let $\mathcal S_{ng}$ be the retained particles assigned to boundary cell $g$
at frame $n$.  If this set is empty, set $Q_{ng}=0$.  Otherwise choose the
particle closest to the wall,
\begin{equation}
  r_{ng}=\mathop{\rm arg\,min}_{i\in\mathcal S_{ng}}d_i^n,
  \qquad
  Q_{ng}=\operatorname{sgn}q_{r_{ng}}^n
  \in\{-1,0,+1\}.
  \label{eq:boundary_frame_representative}
\end{equation}
If two nearest particles are tied within $10^{-7}d_0$ and have opposite signs,
set $Q_{ng}=0$.  Therefore every cell at every frame contributes exactly one
value $Q_{ng}\in\{-1,0,+1\}$, regardless of how many particles occupy that
cell.

\subsection{Checking Heading Sign Against Measured Motion}

The sign of the heading projection must agree with the particle's measured
motion along the boundary.  We track the selected particle $r_{ng}$ from frame
$n$ to frame $n+1$ and calculate
\begin{equation}
  \Delta s_{i}^{n}
  =\operatorname{wrap}_{[-P/2,P/2)}
  \left(s_i^{n+1}-s_i^n\right).
  \label{eq:wrapped_boundary_increment}
\end{equation}
A pair of frames is a valid trial, $T_{ng}=1$, only if the same particle has a
valid projection onto the same smooth boundary segment at both frames, passes
the particle-selection rule at both frames, has the same nonzero sign of $q$ at
both frames, and has a unique wrapped displacement.  On a straight boundary
segment, the largest accepted projected step is
$J=1.05v\Delta t_s$.  On a curved segment, it is
\begin{equation}
  J=
  \frac{1.05v\Delta t_s}{1-\kappa_b d_{\max}},
  \qquad
  d_{\max}=\begin{cases}
  d_0,&\alpha>\pi/2,\\
  \max(d_i^n,d_i^{n+1})+v\Delta t_s,&0<\alpha\le\pi/2.
  \end{cases}
  \label{eq:projection_increment_bound}
\end{equation}
The trial is rejected if the denominator is not positive, if $J\ge P/2$, or if
$|\Delta s_i^n|>J$.  A valid trial is counted as a successful match when
\begin{equation}
  E_{ng}=T_{ng}
  \mathbf 1\left[
  \frac{|\Delta s_{r_{ng}}^n|}{v\Delta t_s}\ge0.02
  \right]
  \mathbf 1\left[
  \operatorname{sgn}\Delta s_{r_{ng}}^n=Q_{ng}
  \right].
  \label{eq:kinematic_calibration_success}
\end{equation}
Clockwise and counterclockwise trials are pooled.  Hence this check does not
choose a preferred chirality.  Boundary cell $g$ passes the check when
\begin{equation}
  \sum_nT_{ng}\ge3,
  \qquad
  C_g=\frac{\sum_nE_{ng}}{\sum_nT_{ng}}\ge\frac{2}{3}.
  \label{eq:kinematic_calibration_gate}
\end{equation}
Thus a cell needs at least three valid trials, and at least two thirds of those
trials must have displacement in the direction predicted by $Q_{ng}$.  Trial
counts do not weight either final quantity.  They only decide whether cell $g$
may contribute.  The calibration and the two final quantities use the same
terminal window.  We therefore also report the fraction of boundary length
that passes this calibration.

\subsection{From Frame Signs to Time-Block Signs}

A directional signal must appear in the same boundary cell in two consecutive
frames.  Define
\begin{equation}
  Y_{ng}=\begin{cases}
  Q_{ng},&Q_{ng}=Q_{n+1,g}\ne0,\\
  0,&\text{otherwise}.
  \end{cases}
  \label{eq:two_frame_phase_support}
\end{equation}
The selected particle at frame $n$ may differ from the selected particle at
frame $n+1$.  Equation~\eqref{eq:two_frame_phase_support} therefore keeps a
boundary signal when one particle replaces another with the same chirality.
No particle is required to travel around the full boundary.

\begin{figure}[H]
\centering
\includegraphics[width=\textwidth]{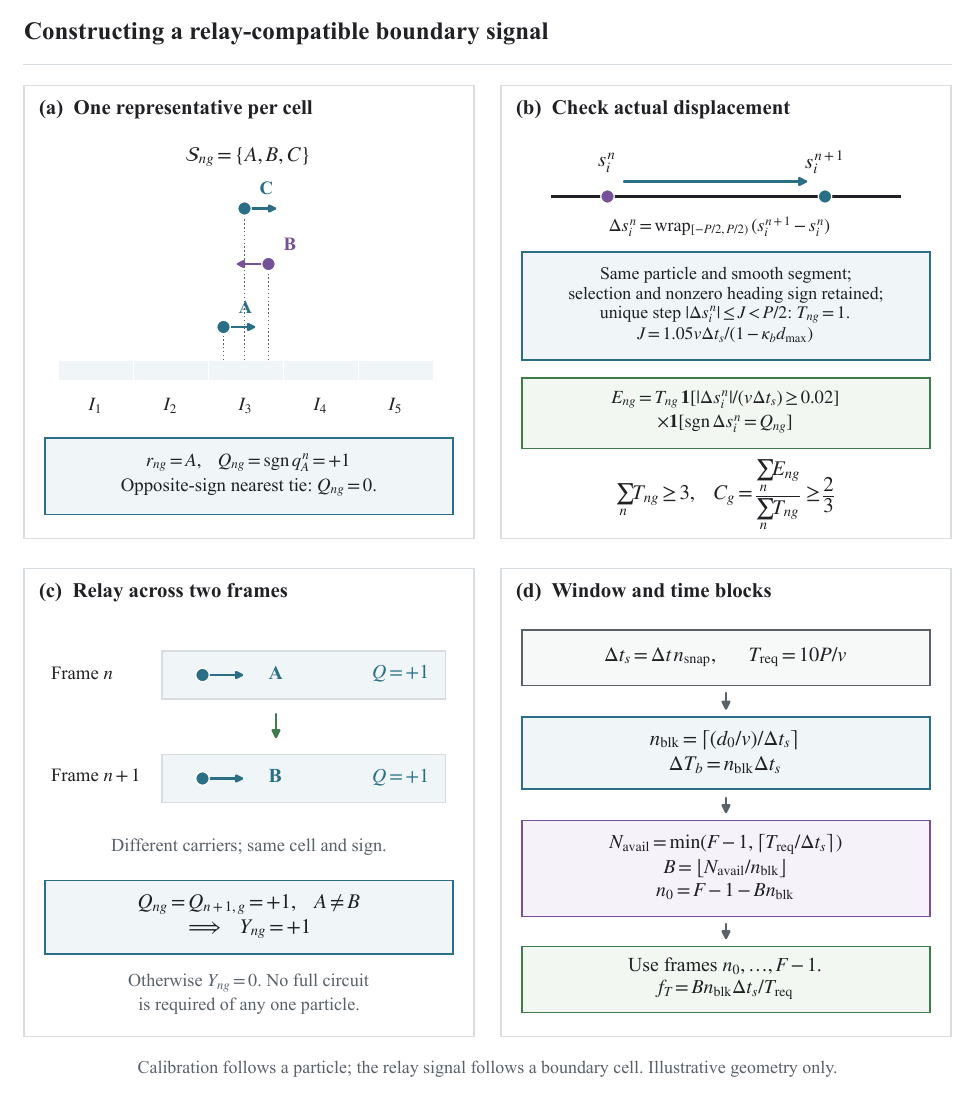}
\caption{Code-drawn construction of the discrete boundary signal.  (a) The
nearest retained particle supplies the single-frame sign $Q_{ng}$ for one
boundary cell.  (b) The heading sign is checked against the measured
arclength displacement through $T_{ng}$, $E_{ng}$, and $C_g$.  This
calibration is the only stage that requires the same particle in consecutive
frames.  (c) Equal nonzero cell signs give $Y_{ng}$ even when the selected
particle changes, thereby retaining relay transport.  (d) The saved-frame count,
saved-frame spacing, perimeter, interaction distance, and speed determine the
terminal observation window and its time-block indices.  The example contains
no simulation data.}
\label{fig:relay_signal_schematic}
\end{figure}

Figure~\ref{fig:relay_signal_schematic} follows one cell from particle
selection to the two-frame relay signal and the independent displacement
check.  The particle labels in the drawing are illustrative.

The target duration of one time block is $d_0/v$.  Because the analysis uses
saved frames, the number of saved intervals per block is
\begin{equation}
  n_{\rm blk}=\left\lceil\frac{d_0/v}{\Delta t_s}\right\rceil,
  \qquad
  \Delta T_b=n_{\rm blk}\Delta t_s.
  \label{eq:boundary_block_time}
\end{equation}
For the present data, $n_{\rm blk}=7$ and $\Delta T_b=0.35$; the continuous
target is $d_0/v=1/3$.  The requested observation time is ten free-flight
circuits of the boundary, $T_{\rm req}=10P/v$.  For a file with $F$ saved
frames, the number of complete blocks and the first used frame are
\begin{equation}
 N_{\rm avail}=\min\!\left(F-1,
 \left\lceil\frac{T_{\rm req}}{\Delta t_s}\right\rceil\right),
 \quad
 B=\left\lfloor\frac{N_{\rm avail}}{n_{\rm blk}}\right\rfloor,
 \quad
 n_0=F-1-Bn_{\rm blk}.
 \label{eq:terminal_window_blocks}
\end{equation}
Only the final $Bn_{\rm blk}$ saved intervals, from frame $n_0$ through frame
$F-1$, are used.  We record
\begin{equation}
 f_T=\frac{Bn_{\rm blk}\Delta t_s}{T_{\rm req}}
 \label{eq:analysis_window_fraction}
\end{equation}
to quantify the retained observation time. Discarding an incomplete final
block can give $f_T<1$ even for a sufficiently long file; the plots use
$f_T\ge0.995$ to label a near-complete requested window.
In block $b$ and cell $g$, count the positive and negative
two-frame signals:
\begin{align}
  N_{bg}^{+}&=\sum_{n\in b}\mathbf 1(Y_{ng}=+1),
  &
  N_{bg}^{-}&=\sum_{n\in b}\mathbf 1(Y_{ng}=-1),
  \label{eq:block_sign_counts}\\
  p_{bg}&=
  \begin{cases}
  \displaystyle
  \frac{|N_{bg}^{+}-N_{bg}^{-}|}{N_{bg}^{+}+N_{bg}^{-}},
  &N_{bg}^{+}+N_{bg}^{-}>0,\\[0.6em]
  0,&N_{bg}^{+}+N_{bg}^{-}=0.
  \end{cases}
  \label{eq:block_conditional_purity}
\end{align}
The directional agreement within this cell and block is $p_{bg}$.  It equals
zero when no two-frame signal is present.  The final sign assigned to this cell
and block is
\begin{equation}
  Z_{bg}=\begin{cases}
  \operatorname{sgn}(N_{bg}^{+}-N_{bg}^{-}),
  &N_{bg}^{+}+N_{bg}^{-}\ge n_{\min},\quad
   p_{bg}\ge2/3,\quad \text{and cell $g$ is calibrated},\\
  0,&\text{otherwise}.
  \end{cases}
  \label{eq:block_gate_sign}
\end{equation}
The main calculation uses $n_{\min}=1$.  Thus $Z_{bg}$ is nonzero only when the
cell is calibrated, at least one two-frame signal occurs, and the normalized
sign difference satisfies $p_{bg}\ge2/3$.  If both signs occur, this last
condition means that the dominant sign accounts for at least $5/6$ of the
signals.  Each boundary cell contributes one value in $\{-1,0,+1\}$,
independent of the number of particles or events in that cell.  We repeat the
full calculation with $n_{\min}=2$ and $3$ to check the effect of requiring
more two-frame signals.

Before calculating the two final quantities, the signal must cover a connected
part of the boundary over the observation window.  Let
\begin{equation}
  \mathcal A=\{g:\exists b,\ |Z_{bg}|=1\}.
  \label{eq:active_gate_union}
\end{equation}
be the set of cells that are active in at least one block.  We require the
largest connected group of these cells to have total length at least $2d_0$:
\begin{equation}
  \max_{\mathcal G\in\operatorname{Conn}(\mathcal A)}
  \sum_{g\in\mathcal G}\ell_g\ge2d_0.
  \label{eq:macroscopic_extent_gate}
\end{equation}
Disconnected cell groups are not added together.  This condition prevents an
isolated active cell from being counted as boundary transport.  The cells in
the connected interval do not need to be active at the same time: different
particles may carry the signal through different parts of the interval during
different blocks.  The same test is also repeated with connectivity broken at
all removed corners and joins.

\subsection{The Two Stability Measures}

If Eq.~\eqref{eq:macroscopic_extent_gate} is satisfied, block $b$ is active
when at least one boundary cell has a nonzero block sign:
\begin{equation}
  M_b=\mathbf 1\{\exists g:\ |Z_{bg}|=1\}.
  \label{eq:active_boundary_block}
\end{equation}
The first measure is the fraction of active blocks,
\begin{equation}
  \boxed{
  \Xi_{\rm Persist}=\frac{1}{B}\sum_{b=1}^{B}M_b
  }.
  \label{eq:xi_persist}
\end{equation}
Thus $\Xi_{\rm Persist}=1$ means that a directional boundary signal is found in
every time block, while $\Xi_{\rm Persist}=0.5$ means that it is found in half
of the blocks.  This measure does not count how many particles carry the
signal, and it is not multiplied by a carrier fraction.

For each active block, the length-weighted mean chirality is
\begin{equation}
  \chi_b=
  \frac{\sum_g\ell_g Z_{bg}}
       {\sum_g\ell_g|Z_{bg}|}
  \in[-1,1].
  \label{eq:block_mean_chirality}
\end{equation}
Here $\chi_b=+1$ means that all active boundary length in block $b$ is
counterclockwise, $\chi_b=-1$ means that it is clockwise, and values near zero
mean that clockwise and counterclockwise active lengths cancel.  The second
measure is
\begin{equation}
  \boxed{
  \Xi_{\rm Sign}=
  \left|
  \frac{\sum_bM_b\chi_b}{\sum_bM_b}
  \right|
  }.
  \label{eq:xi_sign}
\end{equation}
All active blocks, including blocks with $\chi_b=0$, remain in the average.
$\Xi_{\rm Sign}=1$ means that $\chi_b=+1$ in every active block or
$\chi_b=-1$ in every active block.  A value near zero means that opposite
chiralities cancel within blocks, reverse between blocks, or both.  Every active block has
equal weight.  $\Xi_{\rm Sign}$ is reported as unavailable when fewer than
three blocks are active.  Both measures are also unavailable if the connected
length test fails.  Neither measure is weighted by particle number, event
number, crossing rate, or current magnitude.

\begin{figure}[H]
\centering
\includegraphics[width=\textwidth]{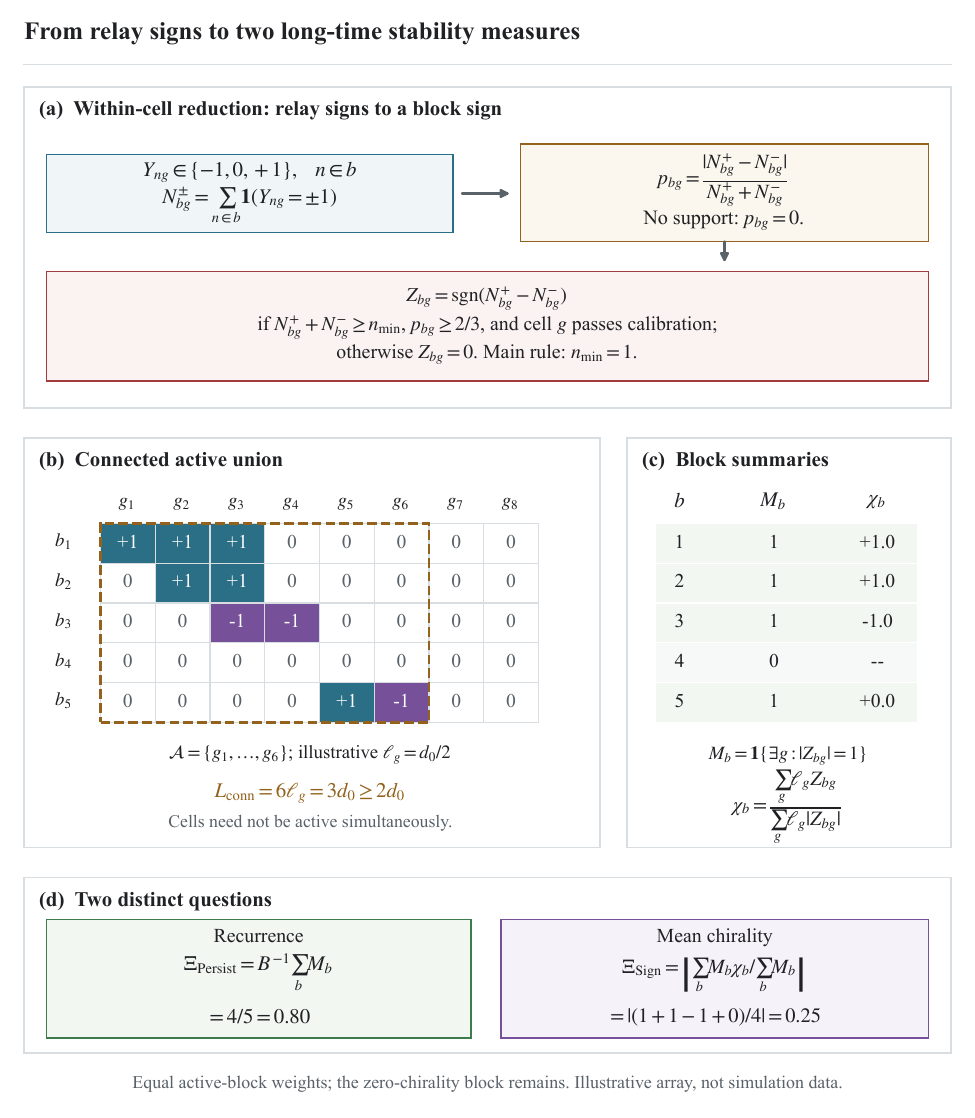}
\caption{Code-drawn numerical example for
$\Xi_{\rm Persist}$ and $\Xi_{\rm Sign}$.  (a) For each cell $g$ and block
$b$, the two-frame signs $Y_{ng}$ give the positive and negative counts,
conditional purity $p_{bg}$, and calibrated block sign $Z_{bg}$.  (b) The
illustrative $Z_{bg}$ array is required to contain a connected active union
$\mathcal A$ of macroscopic length.  (c) Each block is reduced to the activity
indicator $M_b$ and the length-weighted mean chirality $\chi_b$.  (d)
$\Xi_{\rm Persist}$ asks how often a directional signal recurs, whereas
$\Xi_{\rm Sign}$ asks whether the mean chirality keeps one sign.  Neither
quantity uses $W_i^n$ as a numerical weight.  The displayed values are
constructed solely to verify the calculation and are not simulation results.}
\label{fig:stability_measure_schematic}
\end{figure}

Figure~\ref{fig:stability_measure_schematic} gives a complete numerical
example of the reduction from $Z_{bg}$ to the two reported measures.  It shows
that signal recurrence and fixed mean chirality are calculated from the same
cell--block array but answer different questions.

Two geometric quantities are recorded to show how much of the boundary enters
the calculation.  The active boundary-length fraction in block $b$ is
\begin{equation}
  a_b=\frac{1}{P_{\rm reg}}\sum_g\ell_g|Z_{bg}|
  \label{eq:active_arclength_fraction}
\end{equation}
and the fraction of retained boundary length that passes the displacement
check is
\begin{equation}
  f_{\rm cal}=\frac{1}{P_{\rm reg}}
  \sum_g\ell_g\,\mathbf 1[\text{cell $g$ is calibrated}]
  \label{eq:calibrated_perimeter_fraction}
\end{equation}
For $\alpha>\pi/2$, let $O_{ng}^{\rm pre}$ and
$O_{ng}^{\rm sel}$ indicate whether cell $g$ contains at least one candidate
before and after the $W$ requirement, respectively.  The fraction retained by
the $W$ requirement is
\begin{equation}
 \eta_W=
 \begin{cases}
 \displaystyle
 \frac{\sum_{n,g}O_{ng}^{\rm sel}}
      {\sum_{n,g}O_{ng}^{\rm pre}},
 &\sum_{n,g}O_{ng}^{\rm pre}>0,\\[0.8em]
 \mathrm{NA},&\text{otherwise},
 \end{cases}
 \label{eq:w_grid_retention}
\end{equation}
and is reported as unavailable when the $W$ filter is not used.  Wall-distance
quantiles are calculated from the selected nearest particle $r_{ng}$ in each
nonempty cell--frame pair.  The calculation also records the omitted boundary
fraction and the result of the no-cross-join length check.  These quantities
check how the particle selection and geometry affect the two stability
measures; they are not additional measures of transport.

\subsection{Terminal-Window Results and Interpretation}

The analysis uses four geometries and the grid
$\alpha/\pi=0,0.1,\ldots,1$, giving $44$ exact parameter-matched HDF5 files.
For every file, the code checks the required datasets, column numbers,
particle count, and frame alignment. Finite values are checked in the loaded
analysis window, not over each file's entire history; the two phase-lag
endpoints return without loading that window. It then calculates the two
measures for the $36$ rows with $0<\alpha<\pi$, including the four exact
$\alpha=\pi/2$ rows.  No trajectory is restarted or continued.  Every file
uses random seed $9$; therefore the reported values are for one realization of
each parameter set.  Consecutive time blocks are correlated, so no binomial
error bars are assigned.

For $\alpha>\pi/2$, the neighbor sum in
Eq.~\eqref{eq:boundary_free_angular_velocity} is evaluated with a vectorized
particle-pair algorithm.  We compared it with the original simulation loop on
one saved frame from each geometry.  The largest absolute difference was
$1.07\times10^{-14}$; the accepted tolerance was $5\times10^{-10}$.

Thirty-two of the $36$ analyzed rows reach at least $99.5\%$ of the requested
ten-circuit observation time.  Four older circular files are shorter: circular
$0.2\pi$ has $(F-1,B,f_T)=(1000,142,0.678)$, circular $0.4\pi$ has
$(250,35,0.167)$, circular $0.8\pi$ has $(1000,142,0.678)$, and the
single-defect circle at $0.2\pi$ has $(1000,142,0.542)$.  Open markers identify
these four rows; filled markers identify full-window rows.  A shorter window
has fewer chances to contain a rare chirality reversal, even after division by
the number of blocks $B$.

Two details affect how the numbers should be read.  First, the projection mask
removes a length $0.05d_0$ on each side of a corner or defect join.  Second, the
low-$\alpha$ rule has no radial cutoff and can assign a boundary coordinate to
a particle far from the wall.  The reported active-length and wall-distance
values show the size of these effects for every row.

\begin{center}
\scriptsize
\setlength{\tabcolsep}{3.2pt}
\resizebox{\textwidth}{!}{%
\begin{tabular}{c|cc|cc|cc|cc}
\hline
$\alpha/\pi$
& \multicolumn{2}{c|}{Circular}
& \multicolumn{2}{c|}{Circular, $H=3$}
& \multicolumn{2}{c|}{Square}
& \multicolumn{2}{c}{Square, four $H=1$ defects}\\
& $\Xi_{\rm Persist}$ & $\Xi_{\rm Sign}$
& $\Xi_{\rm Persist}$ & $\Xi_{\rm Sign}$
& $\Xi_{\rm Persist}$ & $\Xi_{\rm Sign}$
& $\Xi_{\rm Persist}$ & $\Xi_{\rm Sign}$\\
\hline
0.1 & 1.0000 & 1.0000 & 1.0000 & 0.0360 & 0.9474 & 0.0726 & 0.7055 & 0.0655\\
0.2 & 1.0000 & 0.9862 & 1.0000 & 0.2631 & 0.9774 & 0.0771 & 0.8988 & 0.0611\\
0.3 & 1.0000 & 1.0000 & 0.9962 & 0.0649 & 0.9925 & 0.9879 & 1.0000 & 0.3494\\
0.4 & 1.0000 & 0.9105 & 1.0000 & 0.5694 & 1.0000 & 0.9783 & 1.0000 & 0.4731\\
0.5 & 1.0000 & 1.0000 & 1.0000 & 0.6643 & 1.0000 & 0.9932 & 1.0000 & 0.1531\\
0.6 & 1.0000 & 1.0000 & 1.0000 & 1.0000 & 1.0000 & 1.0000 & 1.0000 & 1.0000\\
0.7 & 1.0000 & 1.0000 & 1.0000 & 1.0000 & 1.0000 & 1.0000 & 1.0000 & 0.9997\\
0.8 & 1.0000 & 1.0000 & 1.0000 & 1.0000 & 1.0000 & 1.0000 & 1.0000 & 0.9998\\
0.9 & 1.0000 & 1.0000 & 1.0000 & 1.0000 & 1.0000 & 1.0000 & 1.0000 & 0.9999\\
\hline
\end{tabular}
}
\end{center}

The machine-readable table also contains
$\overline a=B^{-1}\sum_ba_b$, $f_{\rm cal}$, $\eta_W$, the selected-particle
wall-distance quantiles, $B$, $f_T$, and the HDF5 filename used for every row.
Only $\Xi_{\rm Persist}$ and $\Xi_{\rm Sign}$ are used as boundary-flow
stability measures.

The removed fractions of perimeter are $0$, $0.0109$, $0.0143$, and $0.0467$
for the circle, single-defect circle, square, and four-defect square.  All $36$
rows still pass the connected-length requirement when adjacency is broken at
every removed join.  Thus no row passes only because cells on opposite sides
of a removed join were treated as adjacent.  For every patterned row,
$f_{\rm cal}=1$. The smallest value outside the patterned regime is $0.875000$
for the square at $0.3\pi$; the single-defect circle at $0.1\pi$ gives $0.875520$.

\begin{figure}[htbp]
\centering
\includegraphics[width=\textwidth]{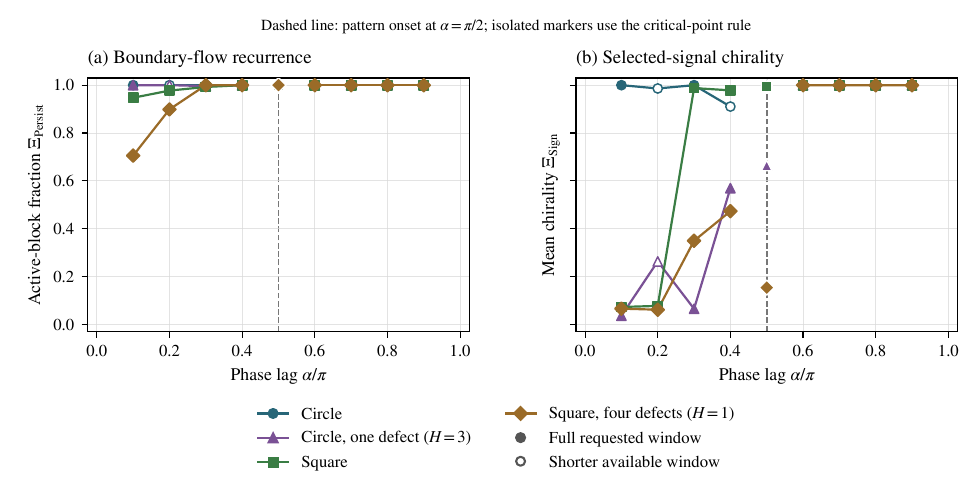}
\caption{The two boundary-flow stability measures on the exact
$0.1\pi$-spaced sweep.  The dashed line marks the pattern-formation threshold
seen in the particle trajectories.  The disconnected $\alpha=\pi/2$ markers
use the critical-point particle-selection rule defined above.  Filled markers
reach the requested ten-circuit window; open markers use the shorter available
record.  The phase-lag endpoints $\alpha=0$ and $\alpha=\pi$ are not assigned these
single-chirality measures.}
\label{fig:phase_informed_boundary_headlines}
\end{figure}

\begin{figure}[htbp]
\centering
\includegraphics[width=\textwidth]{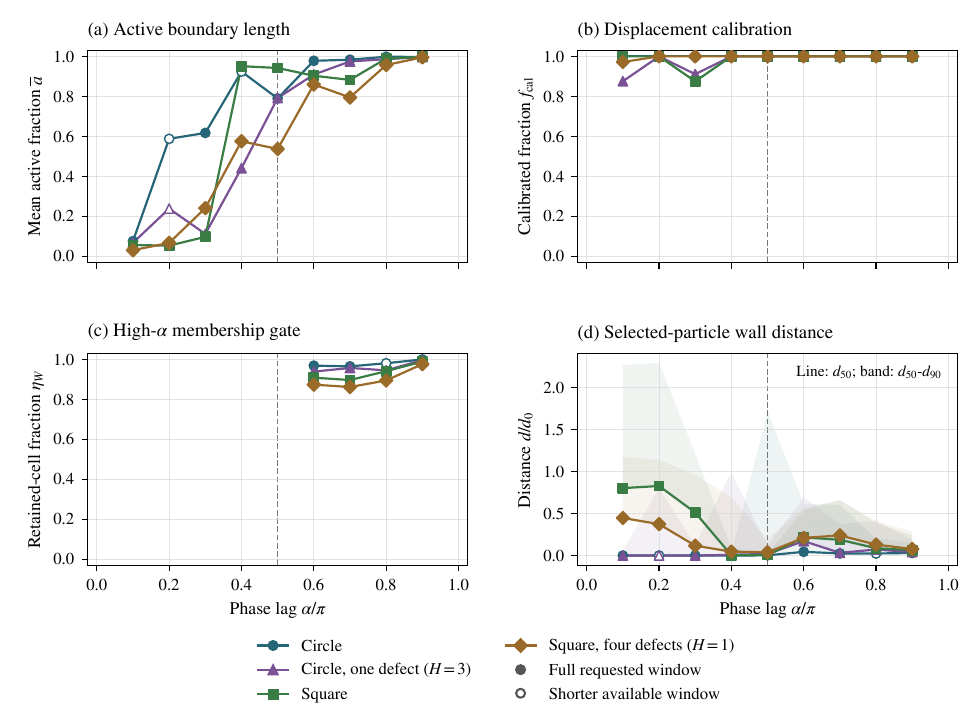}
\caption{Checks on boundary coverage and particle selection.  In the
wall-distance panel, lines show $d_{50}/d_0$ and shaded bands extend to
$d_{90}/d_0$.  Marker fill has the same observation-window meaning as in
Fig.~\ref{fig:phase_informed_boundary_headlines}.}
\label{fig:phase_informed_boundary_diagnostics}
\end{figure}

\begin{figure}[htbp]
\centering
\includegraphics[width=\textwidth]{Figures/publication/circular_boundary_terminal_states.pdf}\\[0.4em]
\includegraphics[width=\textwidth]{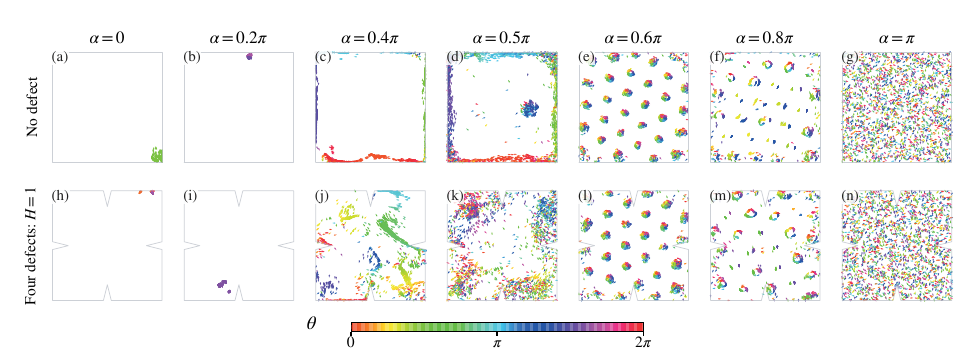}
\caption{Terminal saved states used for direct morphology checks, with heading
phase shown by the common cyclic color map.  The $\alpha=\pi/2$ column displays
the independently classified pattern-formation threshold, while the
$\alpha=\pi$ column is a phase-lag-endpoint morphology comparison and is not included
in the single-chirality measures.  Boundary outlines are deliberately light so that
near-wall particles remain visible.}
\label{fig:terminal_boundary_state_comparison}
\end{figure}

At $0.1\pi$--$0.3\pi$, the square has mean active boundary-length fraction
$\overline a=0.053$--$0.097$, and the four-defect square has
$\overline a=0.030$--$0.241$.  Their selected-particle $d_{90}$ ranges are
$1.208d_0$--$2.294d_0$ and $0.960d_0$--$1.173d_0$, respectively.  Therefore a
low-$\alpha$ signal can appear in many time blocks while occupying only a small
part of the boundary in each block, and some selected particles can be more
than one interaction distance from the wall.  For
$0.6\pi\le\alpha\le0.9\pi$, $\overline a=0.795$--$0.997$ and
$\eta_W=0.862$--$1.000$.  In the patterned regime, the stable chirality is
supported over most of the retained boundary rather than by a few isolated
cell--frame pairs.

For $0.1\pi\le\alpha\le0.4\pi$, the convex circular boundary has
$\Xi_{\rm Persist}=1$ and $\Xi_{\rm Sign}=0.911$--$1.000$.  It therefore
contains a directional boundary signal in every time block, and its mean
chirality remains almost fixed.  With one strong circular defect,
$\Xi_{\rm Persist}=0.996$--$1.000$ but $\Xi_{\rm Sign}$ ranges from $0.036$
to $0.569$.  Thus a boundary signal is still found in nearly every block, but
its mean chirality reverses or cancels over the full window.  The square changes from
$\Xi_{\rm Sign}\simeq0.073$ at $0.1\pi$--$0.2\pi$ to $0.988$ and $0.978$ at
$0.3\pi$ and $0.4\pi$; four symmetric defects retain only $0.065$, $0.061$,
$0.349$, and $0.473$.  The lowest recurrence, $0.706$, occurs for the
four-defect square at $0.1\pi$.  Below the pattern threshold, the regular
circle supports long-time chiral boundary flow, but strong defects and corners
can strongly reduce the stability of its chirality.

For every tested patterned state at
$\alpha/\pi=0.6,0.7,0.8,0.9$, $\Xi_{\rm Persist}=1$.  All
$\Xi_{\rm Sign}$ values are unity to the reported precision except the four-defect square,
which gives $0.9997$, $0.9998$, and $0.9999$ at $0.7\pi$, $0.8\pi$, and
$0.9\pi$.  Thus every tested geometry has a boundary signal in every block and
keeps one mean chirality after the interior pattern has formed.  Direct
trajectory inspection shows the corresponding particle exchange: interior
particles join the boundary flow, while particles scattered by a defect leave
it or rejoin at another boundary segment.  The high-$\alpha$ numbers use the
particle-selection conditions in Eq.~\eqref{eq:high_alpha_candidate_filter};
$\eta_W$ reports retained cell occupancy, not sign-resolved particle counts.
For $\alpha>\pi/2$, these are conditional statistics of the population
selected by Eqs.~\eqref{eq:q0_dead_band} and
\eqref{eq:high_alpha_candidate_filter}, not the net chirality
of all near-wall particles: on a straight segment $W=-\Omega_iq_i$, which
can select one direction. Near-unit $\Xi_{\rm Sign}$ is therefore interpreted
together with direct trajectory observations, rather than as an independent
test of chirality selection.

The isolated threshold $\alpha=\pi/2$ gives $\Xi_{\rm Persist}=1$ for all four
geometries, while $\Xi_{\rm Sign}$ is $1.000$, $0.664$, $0.993$, and $0.153$
for the circular, single-defect circular, square, and four-defect square,
respectively.  These four values are calculated directly from the
$\alpha=\pi/2$ files with the no-radial-cutoff rule and without the $W$ filter.
They are not interpolated from neighboring $\alpha$ values.  At the threshold,
recurrence is high in every geometry, but chirality is still sensitive to the
boundary shape and defects.

The robustness check $q_0=0.02,0.05,0.10$ was performed only for the most
sensitive $\alpha=0.2\pi$ square comparison.  The no-defect values span
$\Xi_{\rm Persist}=0.974$--$0.981$ and
$\Xi_{\rm Sign}=0.077$--$0.084$; the four-defect values span
$0.896$--$0.902$ and $0.061$--$0.074$, respectively.  The small
$\Xi_{\rm Sign}$ values therefore remain when the sign-resolution threshold is
changed by a factor of five.

The minimum supported-pair threshold was also changed from one to two and
three without re-reading or reweighting particles.  The sensitive square
$0.2\pi$ recurrence spans $0.951$--$0.977$ and its net chirality spans
$0.075$--$0.085$; the four-defect values span $0.844$--$0.899$ and
$0.029$--$0.061$.  Requiring more two-frame signals changes the exact
recurrence values but leaves the conclusion unchanged: the long-time mean
chirality is weak for these two cases.

The bulk Chern classification applies where its projector remains isolated;
the finite-strip branches provide a complementary linear comparison.
The particle data add the
following result.  Below $\pi/2$, the regular circle maintains one chirality,
but strong boundary scattering can reverse or cancel that chirality.  Above
$\pi/2$, the interior pattern continuously supplies particles to the boundary,
and both stability measures remain near one for all tested defects and
geometries.  Thus the linear topology gives the chiral boundary mode, while the
interior particle pattern provides the strong protection of the realized
particle flow.  Neither measure contains current magnitude or particle number.
Consequently, their values near one at $0.9\pi$ do not measure the visible
thinning of the boundary layer as $\alpha$ approaches $\pi$.  Every refined
grid point uses an exact HDF5 parameter match; no value is interpolated or
continued from another trajectory.

At the separate singular phase-lag endpoint $\alpha=\pi$, the circular system supports
bidirectional boundary-lattice flow: particles can join, leave, and rejoin
different boundary segments, while interior particles replenish flow scattered
by a circular defect.  The flow does not propagate along the defect boundary;
instead, the interior population supplies the downstream circular-boundary
segments.  The square $\alpha=\pi$ state has no visually resolved boundary flow, with or
without the tested defect, as shown in
Fig.~\ref{fig:terminal_boundary_state_comparison}.  Because these phase-lag-endpoint
states do not contain a single-chirality boundary flow, neither phase-lag endpoint is
assigned $\Xi_{\rm Persist}$ or $\Xi_{\rm Sign}$.

\subsection{Representative Lattice-Scale Comparison}

We compare ten independent realizations (seeds $1$--$10$) at the common
parameters $N=2000$, $2R=7d_0$, $K=20.75$, and $v=3$.  No cluster-number
observable is used.  For a boundary density $\rho_f(\varphi,t)$, where $f$
denotes the full boundary layer or one phase-resolved stream, define
\begin{align}
 \delta\rho_f(\varphi,t)
 &=\rho_f(\varphi,t)-\frac{1}{2\pi}
   \int_0^{2\pi}\rho_f(\varphi',t)\,d\varphi',\\
 C_f(\Delta s,t)
 &=\frac{\int_0^{2\pi}
   \delta\rho_f(\varphi,t)
   \delta\rho_f(\varphi+\Delta s/R_f(t),t)\,d\varphi}
 {\int_0^{2\pi}\delta\rho_f^2(\varphi,t)\,d\varphi},
 \label{eq:boundary_density_correlation}
\end{align}
with $R_f(t)$ the mean radial position of the selected particles in that frame.
The implementation bins the density into $2048$ angular bins, smooths it
with a periodic Gaussian of width $2.5$ bins, computes and zero-lag-normalizes
the circular autocorrelation, and smooths the correlation with width $2.0$ bins.
The first-neighbor peak in each frame is selected by largest prominence in
$0.60d_0\leq\Delta s\leq1.80d_0$, with minimum peak separation $0.35d_0$
and prominence at least the larger of $0.01$ and $0.05$ times the correlation
range within that search window, followed by three-point parabolic interpolation.
The reported spacing is the equal-frame mean of these peak positions, and
its error bar is their temporal sample standard deviation (denominator equal
to the number of sampled frames minus one);
it is not the peak of a time-averaged correlation. A numerical value is
reported only when the mean normalized peak height is at least $0.30$ and the
temporal coefficient of variation of the peak positions is at most $0.10$; otherwise that realization
is retained in the figure as unresolved.  The resolved/unresolved assignments
are unchanged when the two cutoffs are varied over $0.25$--$0.35$ and
$0.08$--$0.12$, respectively.  To prevent panel-dependent magnification, all
three comparison figures use the same vertical-axis span, $0.50d_0$, rather
than data-dependent limits.  For a regular continuum spectrum we use
\begin{equation}
  k_\ast=\underset{k>0}{\operatorname{arg\,max}}
  \max_n\operatorname{Re}\sigma_n(k) .
  \label{eq:lattice_scale_kstar}
\end{equation}

\begin{figure}[H]
\centering
\includegraphics[width=0.80\textwidth]{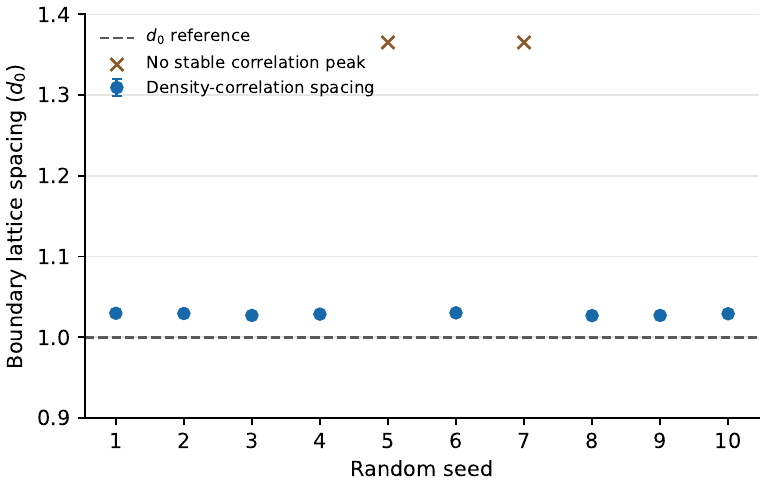}
\caption{Seed-resolved boundary-lattice spacing at $\alpha=\pi/2$.
Error bars are temporal standard deviations over iterations
$40000$--$50000$.  Crosses indicate realizations without a stable correlation
peak; their vertical placement is categorical and has no numerical ordinate.
The dashed $d_0$ line is a length reference, not a linear-spectral prediction.}
\label{fig:halfpi_lattice_seed_robustness}
\end{figure}

At $\alpha=\pi/2$, $a(k)=0$ in Eq.~\eqref{eq:compact_matrix} and all three
growth rates have zero real part for every $k$.  Hence Eq.~\eqref{eq:lattice_scale_kstar}
does not define a spectral prediction at the critical point.  Applying
Eq.~\eqref{eq:boundary_density_correlation} to particles within $0.25d_0$ of
the wall over iterations $40000$--$50000$, sampled every $100$ iterations
($101$ frames), gives
$a_\partial=1.0268$--$1.0302d_0$ (mean $1.0285d_0$) whenever a stable peak is
resolved.  Seeds $5$ and $7$ do not have such a peak and are shown without
assigning a lattice constant.  The $d_0$ line in
Fig.~\ref{fig:halfpi_lattice_seed_robustness} is therefore only the microscopic
interaction-distance reference, not a theoretical prediction.

At $\alpha=0.6\pi$, Eq.~\eqref{eq:lattice_scale_kstar} gives
$k_\ast=5.0910/d_0$.  For a triangular Bravais lattice the first reciprocal
shell satisfies $|\mathbf G_1|=4\pi/(\sqrt{3}a_{\rm hex})$; identifying
$|\mathbf G_1|$ with $k_\ast$ therefore predicts
\begin{equation}
  a_{\rm hex}^{\rm lin}=\frac{4\pi}{\sqrt{3}k_\ast}=1.4251d_0 .
  \label{eq:hex_lattice_spectral_prediction}
\end{equation}
Direct first-shell distances between neighboring vortex centers over iterations
$40000$--$50000$ give seed means $1.3049$--$1.4102d_0$, with ten-seed mean
$1.3568d_0$, or $4.8\%$ below
Eq.~\eqref{eq:hex_lattice_spectral_prediction}.
The centers are estimated from the free angular rates $\Omega_i$ as
$(x_i-v\sin\theta_i/\Omega_i,\,y_i+v\cos\theta_i/\Omega_i)$,
retaining $|\Omega_i|\ge0.2K|\sin\alpha|$ and center radii at most
$R+0.2d_0$. DBSCAN uses neighborhood radius $v/(K|\sin\alpha|)$ and
minimum sample count $10$; retained clusters contain at least $15$ particles
and are represented by coordinate-wise median centers. Centers within
$0.5d_0$ of the wall are excluded. Each analyzed frame is required to contain
at least seven retained centers. First-shell pairs are all unordered pairs separated by at
most $1.45$ times that frame's median nearest-neighbor distance. Pair distances
are averaged within each frame, then equally across the $101$ frames sampled
every $100$ iterations;
error bars are temporal sample standard deviations, not standard errors.

\begin{figure}[H]
\centering
\includegraphics[width=0.80\textwidth]{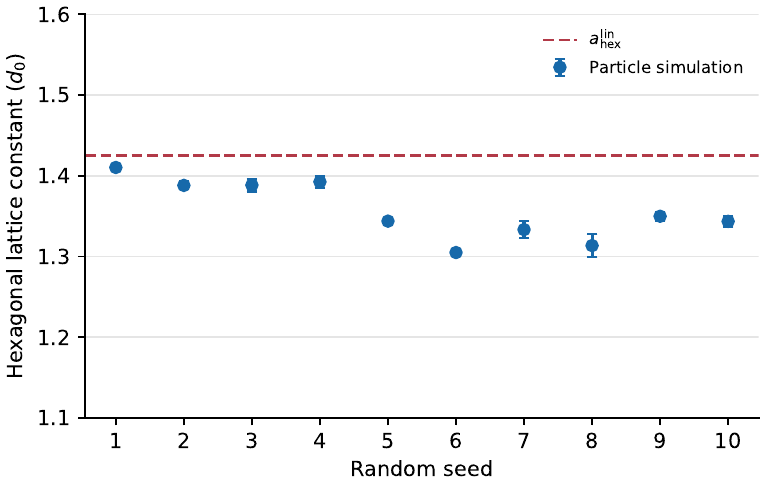}
\caption{Seed-resolved bulk hexagonal lattice constant at
$\alpha=0.6\pi$.  Points are first-shell vortex-center distances, error bars
are their temporal standard deviations, and the dashed line is
Eq.~\eqref{eq:hex_lattice_spectral_prediction}.}
\label{fig:alpha06_hex_lattice_seed_robustness}
\end{figure}

At $\alpha=\pi$, $D_0=0$ and $M(\mathbf k)$ is undefined, so no endpoint
spectrum is assigned.  A candidate scale can only be obtained from the left.
Writing $x=kd_0$, the limiting real-growth envelope is proportional to
$-J_1(x)/x$; its first positive maximum satisfies $J_2(x)=0$.  Thus, with $j_{2,1}$
the first positive zero of $J_2$,
\begin{equation}
  \begin{aligned}
  k_\ast^-d_0
  &=\lim_{\alpha\to\pi^-}k_\ast d_0=j_{2,1}=5.1356,\\
  \ell_\pi^-&\equiv\frac{2\pi}{k_\ast^-}=1.2235d_0,
  \qquad
  a_{\pi}^{-,\triangle}\equiv\frac{2}{\sqrt{3}}\ell_\pi^-
  =\frac{4\pi}{\sqrt{3}k_\ast^-}=1.4127d_0 .
  \end{aligned}
  \label{eq:pi_left_lattice_scale}
\end{equation}
\emph{Preliminary conjecture (not a demonstrated mechanism).---}
Periodic-boundary trajectories at $\alpha=\pi$ suggest a dynamic state formed
by two hexagonal cluster lattices translating in opposite spatial directions.
We conjecture that a predominantly circular collision boundary filters the
length scales carried by this moving bulk structure and exposes a nearby
hexagonal scale as a tangential boundary period.  Under this conjecture,
$\ell_\pi^-$ is interpreted as a local triangular-plane spacing and the
measured period as the associated nearest-neighbor distance, which motivates
the factor $2/\sqrt{3}$ in Eq.~\eqref{eq:pi_left_lattice_scale}.  The morphology
comparison in Fig.~\ref{fig:terminal_boundary_state_comparison} is consistent
with, but does not prove, this picture: the bidirectional lattice is resolved
when most of the wall remains circular, survives one isolated localized
defect, and is not resolved for the tested square boundary.  Thus
$a_{\pi}^{-,\triangle}$ is only a conditional geometric candidate, not a
separate endpoint spectrum or an established consequence of the boundary.
For the particle data, the two counterpropagating populations are first
separated by the two axial families of the relative phase
$\theta_i-\varphi_i$ and are then labelled CCW for the family with the larger
mean tangential heading and CW for the other. In each frame the axial direction is
$\frac12\arg\langle e^{2i(\theta_i-\varphi_i)}\rangle$, evaluated in the
$0.50d_0$ wall shell; particles are retained when the absolute cosine of
their relative phase measured from this axis is at least $0.5$, and the
sign of that projection assigns the two families.
Applying Eq.~\eqref{eq:boundary_density_correlation}
separately to the two families in the $0.50d_0$ wall shell over the last ten
free-flight circuit times (iterations $35300$--$50000$, every $50$ iterations,
$295$ frames) gives resolved seed--stream values
$1.2667$--$1.5099d_0$, with mean $1.3491d_0$.  The mean is $10.3\%$ above
$\ell_\pi^-$ but only $4.5\%$ below the geometry-corrected
$a_{\pi}^{-,\triangle}$.  The CW family of seed $4$ has no stable peak and is
retained as unresolved.

\begin{figure}[H]
\centering
\includegraphics[width=0.80\textwidth]{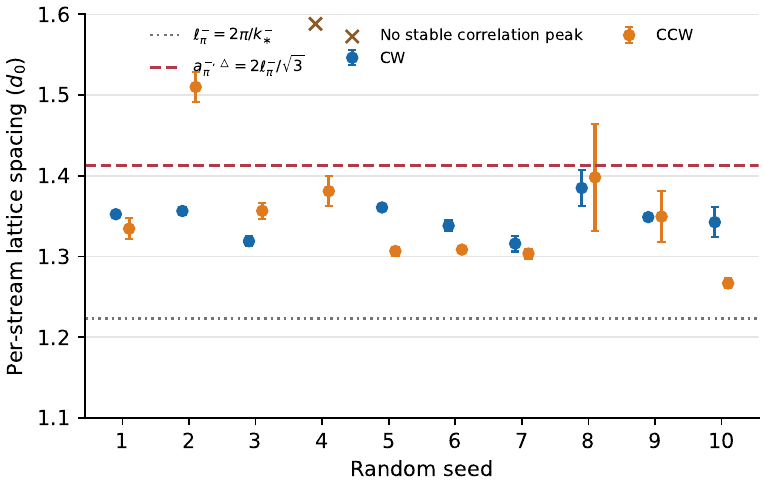}
\caption{Seed- and stream-resolved boundary-lattice spacing at
$\alpha=\pi$.  Error bars are temporal standard deviations.  The cross marks
an unresolved correlation peak and has no numerical ordinate.  The dashed line
is the geometry-corrected left-limit candidate $a_{\pi}^{-,\triangle}$, while
the dotted line is the unconverted $\ell_\pi^-$.  Neither is a spectrum
evaluated at $\alpha=\pi$.}
\label{fig:pi_lattice_seed_robustness}
\end{figure}

The ten-seed comparison separates wavelength selection from state formation.
At $\alpha=\pi/2$, a resolved boundary lattice has a local spacing robustly
close to $d_0$, although some random initial conditions do not reach a state
with a stationary correlation peak.  At $0.6\pi$, the geometry-corrected
linear scale captures the bulk hexagonal spacing to within $4.8\%$.  At
$\alpha=\pi$, applying the triangular-plane conversion to the left-limit scale
reduces the mean discrepancy from $10.3\%$ to $4.5\%$, although the
seed-to-seed spread remains appreciably larger.  This numerical agreement
supports the preliminary boundary-filtering conjecture but cannot distinguish
it from curvature-induced mode locking or finite-circumference commensuration.
Thus the spectral peak selects a preferred scale where it is defined, while
nonlinear relaxation, confinement, and initial-condition dependence determine
whether and how sharply that scale is realized.

\section{Spectral Projectors and Chern Numbers}

In this section bold $\mathbf k=(k_x,k_y)$ is the spatial wave vector and
$k=|\mathbf k|$. We assume $D_0\ne0$, so that the closed matrix
$M(\mathbf k)$ is finite. We also fix a target spectral cluster
$\mathcal I$ whose eigenvalues remain separated from the complementary
spectrum. Separation may be supplied by a closed contour or by a line gap
that keeps the same set of eigenvalues on one side of the line. The Chern
number belongs to the resulting spectral projector, not to the unordered
set of eigenvalues alone. If $D_0=0$ or the separating gap closes, the
projector used below and its Chern number are not defined.

\subsection{Riesz Projector and the Projector Bundle}

For each $\mathbf k$, choose a positively oriented contour
$\Gamma_{\mathbf k}\subset\mathbb C$ that encloses precisely the
eigenvalues in $\mathcal I$ and does not intersect
$\operatorname{Spec}M(\mathbf k)$. The associated Riesz projector is
\begin{equation}
  P_{\mathcal I}(\mathbf k)=\frac{1}{2\pi i}
  \oint_{\Gamma_{\mathbf k}}
  (zI_3-M(\mathbf k))^{-1}\,dz .
  \label{eq:riesz_projection}
\end{equation}
The result is independent of any deformation of the contour that does not
cross the spectrum. On each sufficiently small parameter patch one may
choose a single fixed contour $\Gamma$; projectors obtained from different
local contours agree on overlaps because they enclose the same spectral
cluster.

The resolvent identity
\begin{equation}
  (zI_3-M)^{-1}(wI_3-M)^{-1}
  =\frac{(zI_3-M)^{-1}-(wI_3-M)^{-1}}{w-z}
\end{equation}
inserted into the double contour integral gives
\begin{equation}
  P_{\mathcal I}^2=P_{\mathcal I}.
  \label{eq:projector_idempotency}
\end{equation}
Moreover, differentiating the resolvent on a local parameter patch gives
\begin{equation}
  dP_{\mathcal I}=\frac{1}{2\pi i}
  \oint_{\Gamma}
  (zI_3-M)^{-1}(dM)(zI_3-M)^{-1}\,dz .
  \label{eq:riesz_derivative}
\end{equation}
Thus $P_{\mathcal I}$ is smooth whenever $M$ is smooth and the spectral
separation persists. Its rank
$r=\operatorname{rank}P_{\mathcal I}=\operatorname{Tr}P_{\mathcal I}$ is
then constant, and
\begin{equation}
  E_{\mathcal I}=\left\{
  (\mathbf k,u):P_{\mathcal I}(\mathbf k)u=u
  \right\}
  \longrightarrow\mathcal M
  \label{eq:projector_bundle}
\end{equation}
is a rank-$r$ complex vector bundle over the parameter manifold
$\mathcal M$. Internal degeneracies inside $\mathcal I$ do not invalidate
this construction; only contact between $\mathcal I$ and its complement
does. A globally isolated single band is the special case $r=1$.

\subsection{Biorthogonal Frames and Gauge Freedom}

For a simple eigenvalue $\sigma_n$, define right and left eigenvectors by
\begin{equation}
  M|u_n^R\rangle=\sigma_n|u_n^R\rangle,
  \qquad
  M^\dagger|u_n^L\rangle=\sigma_n^*|u_n^L\rangle,
  \label{eq:left_right_eigenvectors}
\end{equation}
or, equivalently,
$\langle u_n^L|M=\sigma_n\langle u_n^L|$. For a rank-$r$ cluster, choose a
smooth local frame of $\operatorname{Im}P_{\mathcal I}$ and its dual frame
in $\operatorname{Im}P_{\mathcal I}^\dagger$:
\begin{equation}
  R_{\mathcal I}=
  (|r_1\rangle,\ldots,|r_r\rangle),
  \qquad
  L_{\mathcal I}=
  (|\ell_1\rangle,\ldots,|\ell_r\rangle),
\end{equation}
where $R_{\mathcal I},L_{\mathcal I}\in\mathbb C^{3\times r}$. They are
chosen biorthonormally,
\begin{equation}
  L_{\mathcal I}^\dagger R_{\mathcal I}=I_r,
  \label{eq:biorthonormal_frame}
\end{equation}
so that the Riesz projector has the frame representation
\begin{equation}
  P_{\mathcal I}=R_{\mathcal I}L_{\mathcal I}^\dagger,
  \qquad
  P_{\mathcal I}^2=P_{\mathcal I}.
  \label{eq:projector_frame}
\end{equation}
Unlike a Hermitian orthogonal projector, $P_{\mathcal I}^\dagger$ need not
equal $P_{\mathcal I}$. The left frame is therefore essential. When the
cluster is diagonalizable, $|r_a\rangle$ and $|\ell_a\rangle$ may be
chosen as its right and left eigenvectors. At an internal exceptional
degeneracy the individual eigenvectors need not form a basis, but the
smooth Riesz subspace and the frames used in
\eqref{eq:projector_frame} remain well defined as long as the cluster stays
separated from its complement.

The local frame is not unique. For any smooth
$G(\mathbf k)\in GL(r,\mathbb C)$,
\begin{equation}
  R_{\mathcal I}\mapsto R_{\mathcal I}G,
  \qquad
  L_{\mathcal I}\mapsto
  L_{\mathcal I}(G^{-1})^\dagger .
  \label{eq:frame_gauge_transformation}
\end{equation}
This transformation preserves both
$L_{\mathcal I}^\dagger R_{\mathcal I}=I_r$ and
$P_{\mathcal I}=R_{\mathcal I}L_{\mathcal I}^\dagger$. Consequently, a
topological invariant must be independent of this nonunitary frame
freedom.

\subsection{Connection, Curvature, and the First Chern Number}

On a local patch define the non-Abelian biorthogonal connection and its
curvature by
\begin{equation}
  \mathcal A=L_{\mathcal I}^\dagger dR_{\mathcal I},
  \qquad
  \mathcal F=d\mathcal A+\mathcal A\wedge\mathcal A .
  \label{eq:connection_curvature}
\end{equation}
Here $d$ is the exterior derivative on $\mathcal M$; no extra factor of
$i$ is included in this mathematical convention. Under
\eqref{eq:frame_gauge_transformation},
\begin{equation}
  \mathcal A\mapsto G^{-1}\mathcal A G+G^{-1}dG,
  \qquad
  \mathcal F\mapsto G^{-1}\mathcal F G .
  \label{eq:connection_gauge_law}
\end{equation}
Therefore $\operatorname{Tr}\mathcal F$ is frame independent.

The frame and projector forms of the curvature are equivalent. From
$d(L_{\mathcal I}^\dagger R_{\mathcal I})=0$ and
$dP_{\mathcal I}=dR_{\mathcal I}L_{\mathcal I}^\dagger+
R_{\mathcal I}dL_{\mathcal I}^\dagger$, one obtains
\begin{align}
  \operatorname{Tr}(P_{\mathcal I}dP_{\mathcal I}\wedge dP_{\mathcal I})
  &=\operatorname{Tr}\!\left[
  dL_{\mathcal I}^\dagger(I_3-P_{\mathcal I})\wedge
  dR_{\mathcal I}\right]\nonumber\\
  &=\operatorname{Tr}(d\mathcal A+
  \mathcal A\wedge\mathcal A)
  =\operatorname{Tr}\mathcal F .
  \label{eq:projector_frame_curvature}
\end{align}
The identity $P_{\mathcal I}^2=P_{\mathcal I}$ also implies
\begin{equation}
  P_{\mathcal I}dP_{\mathcal I}+(dP_{\mathcal I})P_{\mathcal I}
  =dP_{\mathcal I},
  \qquad
  P_{\mathcal I}(dP_{\mathcal I})P_{\mathcal I}=0 .
  \label{eq:projector_differential_identities}
\end{equation}
Thus first-order changes of the projector connect the target subspace to
its complement.

The first Chern number is the Chern--Weil integral
\begin{equation}
  C(P_{\mathcal I})=\frac{1}{2\pi i}
  \int_{\mathcal M}\operatorname{Tr}\mathcal F
  =\frac{1}{2\pi i}
  \int_{\mathcal M}
  \operatorname{Tr}(P_{\mathcal I}dP_{\mathcal I}\wedge dP_{\mathcal I}).
  \label{eq:projector_chern}
\end{equation}
In Cartesian coordinates this becomes
\begin{equation}
  C(P_{\mathcal I})=\frac{1}{2\pi i}
  \int_{\mathcal M}
  \operatorname{Tr}\!\left\{
  P_{\mathcal I}
  [\partial_{k_x}P_{\mathcal I},\partial_{k_y}P_{\mathcal I}]
  \right\}\,dk_xdk_y .
  \label{eq:projector_chern_cartesian}
\end{equation}

For completeness, its integrality follows directly from the frame gauge
law. Cover a compact $\mathcal M$ by two patches whose frames are related
on their overlap by $G$. Taking the trace of
\eqref{eq:connection_gauge_law} gives
\begin{equation}
  \operatorname{Tr}\mathcal A_2-
  \operatorname{Tr}\mathcal A_1=d\log\det G .
\end{equation}
Stokes' theorem then reduces \eqref{eq:projector_chern} to the transition
function winding,
\begin{equation}
  C(P_{\mathcal I})=
  \frac{1}{2\pi i}\oint d\log\det G\in\mathbb Z,
  \label{eq:chern_integrality}
\end{equation}
up to the orientation assigned to the overlap curve. For $r>1$, this is
the first Chern number of the determinant line bundle
$\det E_{\mathcal I}$; for $r=1$, it is the usual biorthogonal Berry
bundle.

\subsection{Biorthogonal Fukui Discretization}

Let $\mathbf k_{ij}$ be an oriented grid on $\mathcal M$ and let
$R_{ij},L_{ij}\in\mathbb C^{3\times r}$ be local frames satisfying
$L_{ij}^\dagger R_{ij}=I_r$. Define the nearest-neighbor overlap matrices
\begin{equation}
  W_x(i,j)=L_{ij}^\dagger R_{i+1,j},
  \qquad
  W_y(i,j)=L_{ij}^\dagger R_{i,j+1}.
  \label{eq:fukui_overlap}
\end{equation}
Provided their determinants do not vanish, the determinant-line link
variables are
\begin{equation}
  U_x(i,j)=\frac{\det W_x(i,j)}{|\det W_x(i,j)|},
  \qquad
  U_y(i,j)=\frac{\det W_y(i,j)}{|\det W_y(i,j)|}.
  \label{eq:fukui_links}
\end{equation}
The oriented plaquette phase and the discrete Chern number are
\begin{align}
  F_{xy}(i,j)&=\operatorname{Arg}\!\left[
  U_x(i,j)U_y(i+1,j)
  U_x(i,j+1)^{-1}U_y(i,j)^{-1}
  \right],\\
  C_{\rm Fukui}&=\frac{1}{2\pi}\sum_{i,j}F_{xy}(i,j).
  \label{eq:fukui_chern}
\end{align}
Indeed, for a grid spacing $\Delta k_\mu$,
\begin{equation}
  W_\mu=I_r+\mathcal A_\mu\Delta k_\mu+O(\Delta k_\mu^2),
  \qquad
  \det W_\mu=
  \exp\!\left[
  \operatorname{Tr}\mathcal A_\mu\Delta k_\mu+O(\Delta k_\mu^2)
  \right].
  \label{eq:fukui_continuum_limit}
\end{equation}
Before normalization, the product of the four determinant overlaps around
a plaquette satisfies
\begin{equation}
  \widetilde{\mathcal U}_{\square}
  =\exp\!\left[
  \int_{\square}\operatorname{Tr}\mathcal F+O(\Delta k^3)
  \right].
  \label{eq:fukui_plaquette_limit}
\end{equation}
For a non-Hermitian frame this quantity need not have unit modulus.
Equation \eqref{eq:fukui_links} retains its phase: the discarded positive
factor is topologically contractible because
$\mathbb C^\times\simeq\mathbb R_+\times U(1)$, whereas the $U(1)$ factor
carries the winding. Under the frame change
\eqref{eq:frame_gauge_transformation},
\begin{equation}
  W_{ab}\mapsto G(a)^{-1}W_{ab}G(b),
\end{equation}
so all determinant gauge factors cancel around a closed plaquette. The
algorithm is consequently frame independent. On a sufficiently fine
admissible grid, for which no plaquette flux crosses the principal-branch
cut and no $\det W_\mu$ vanishes,
\begin{equation}
  C_{\rm Fukui}\longrightarrow C(P_{\mathcal I}).
\end{equation}
The determinant is essential for a band cluster because it discretizes
the connection on $\det E_{\mathcal I}$ rather than selecting an arbitrary
eigenvector inside the cluster.

The implementation first screens target--complement separation on a radial
grid independent of the compactification mesh, augmented by zeros of $b(k)$
and refined roots of the cubic discriminant. It uses $1201$ logarithmic and
$1201$ linear radial samples before refinement, with gap tolerances
$10^{-6}$ relative and $10^{-9}$ absolute; an operator-norm bound controls
the remaining large-$k$ tail. Passing this finite-radius screening is
numerical evidence, not a proof over every finite $k$.
Frames are constructed by an ordered complex Schur decomposition and a
Sylvester equation for the Riesz projector, so an internal cluster
degeneracy need not be resolved into individual eigenvectors. Rotational
covariance supplies the angular frames. The cap basis is chosen from the
tracked asymptotic spin sectors, and an explicitly supplied incompatible
cap is rejected. Unresolved determinant links, principal-phase ambiguity,
or failure of the independent pole-formula check yield an unavailable
result, rather than dropping plaquettes and rounding the remaining flux.
The main-text platform plots use the screened pole formula, with
representative independent determinant-link evaluations.

Parameter scans return an unavailable value, rather than zero, when the
closure is singular or the target projector is not isolated from its
complement. Such samples interrupt the plotted curve and are retained with
an invalidity reason in the numerical output. In particular, neither
$\alpha=0$ nor $\alpha=\pi$ is labelled a trivial Chern phase at $\omega=0$.

\section{Rotational Covariance and the Origin of \texorpdfstring{$C=\pm2$}{C = +/-2}}

\subsection{Covariant Form and Radial Spectrum}

Write
\begin{equation}
  k_x=k\cos\phi,\qquad k_y=k\sin\phi.
\end{equation}
Define
\begin{equation}
  S(\phi)=
  \begin{pmatrix}
  1&0&0\\
  0&\cos\phi&-\sin\phi\\
  0&\sin\phi&\cos\phi
  \end{pmatrix}.
\end{equation}
The corresponding generator on density--polarization space is
\begin{equation}
  J=iS(\phi)^{-1}\partial_\phi S(\phi)=
  \begin{pmatrix}
  0&0&0\\
  0&0&-i\\
  0&i&0
  \end{pmatrix}.
  \label{eq:angular_generator}
\end{equation}
This $J$ is distinct from the microscopic coupling strength $K$ in
\eqref{eq:particle_coupling_appendix}.
Then
\begin{equation}
  M(k,\phi)=S(\phi)B(k)S(\phi)^{-1},
  \label{eq:rotational_covariance}
\end{equation}
with
\begin{equation}
  B(k)=
  \begin{pmatrix}
  0&-ivk&0\\[4pt]
  -\dfrac{iv}{2}k&a(k)&b(k)\\[4pt]
  0&-b(k)&a(k)
  \end{pmatrix}.
  \label{eq:radial_matrix}
\end{equation}
The covariance relation implies directional spectral isotropy. For the
characteristic polynomial,
\begin{align}
  \chi_M(\sigma;k,\phi)
  &=
  \det[M(k,\phi)-\sigma I_3]\nonumber\\
  &=
  \det[S(\phi)(B(k)-\sigma I_3)S(\phi)^{-1}]\nonumber\\
  &=
  \det[B(k)-\sigma I_3].
  \label{eq:direction_independent_characteristic}
\end{align}
Thus, away from degeneracies, every continuously labelled branch satisfies
$\sigma_n(k,\phi)=\sigma_n(k)$; at a degeneracy the same statement applies
to the isolated spectral cluster. The angular coordinate does not move the
spectrum in the complex plane; it only rotates the eigenvectors and spectral
subspaces. More precisely, let
\begin{equation}
  P_{\mathcal I}^{(0)}(k):=P_{\mathcal I}(k,0)
\end{equation}
be the Riesz projector of the radial matrix $B(k)$. Since
\begin{equation}
  (zI_3-M(k,\phi))^{-1}
  =S(\phi)(zI_3-B(k))^{-1}S(\phi)^{-1},
\end{equation}
the full two-dimensional projector is
\begin{equation}
  P_{\mathcal I}(k,\phi)=
  S(\phi)P_{\mathcal I}^{(0)}(k)S(\phi)^{-1}.
  \label{eq:projector_rotational_covariance}
\end{equation}
The superscript $(0)$ denotes the reference direction $\phi=0$; it is not
a band label.

\subsection{Finite-Wavenumber Isolation and Exceptional Points}

The condition $D_0\ne0$ makes the closed matrix finite but does not guarantee
an isolated spectral projector. This must be checked along the entire radial
spectrum, whose characteristic equation is
\begin{equation}
 \det[\sigma I_3-B(k)]
 =\sigma^3-2a\sigma^2+
 (a^2+b^2+v^2k^2/2)\sigma-av^2k^2/2=0.
 \label{eq:radial_characteristic_polynomial}
\end{equation}
For example, with $v=3$, $\omega=0$, $d_0=1$,
$K=\lambda\rho_0\widehat G(0)=20.75$, and $\alpha=0.99\pi$, the oscillatory pair
meets at finite-wavenumber exceptional points near $k=0.413267$ and
$0.472217$. Its separate rank-one Chern numbers are therefore undefined,
despite $D_0\ne0$ and distinct pole sectors. An exceptional point internal
to a band cluster does not by itself invalidate that cluster: the relevant
condition is separation of the target cluster from its complement.
Consequently the open phase-lag interval alone is not a Chern phase diagram.

\subsection{Compactification of the Projector Base Space}

The punctured wave-number plane is $(0,\infty)\times S^1$. In polar
coordinates, the circle at $k=0$ represents the single origin, while the
one-point compactification also collapses the circle at $k=\infty$.
Collapsing both boundary circles produces
\begin{equation}
  \mathbb R^2\cup\{\infty\}\simeq S^2.
  \label{eq:wavenumber_compactification}
\end{equation}
For a spectral bundle, this topological compactification is legitimate
only if the chosen projector has direction-independent limits at both
ends:
\begin{equation}
  \lim_{k\to0}P_{\mathcal I}(k,\phi)=P_{\mathcal I}(0),
  \qquad
  \lim_{k\to\infty}P_{\mathcal I}(k,\phi)=P_{\mathcal I}(\infty),
  \label{eq:projector_pole_limits}
\end{equation}
uniformly in $\phi$. Directional spectral isotropy is necessary for using
one radial spectral separation at every angle, but it is not by itself
sufficient; the polar projector limits in
\eqref{eq:projector_pole_limits} are the additional requirement.

At $k=0$, $M(0,\phi)=B(0)$ is independent of $\phi$ and commutes with
$S(\phi)$. Its isolated Riesz projectors therefore commute with
$S(\phi)$ as well. At large $k$, the disk kernel obeys
$\widehat G(k)\to0$, and
\begin{equation}
  \frac{B(k)}{k^2}\longrightarrow
  A_\infty=i\beta J,
  \qquad
  \beta=\frac{v^2}{4D_0},
  \label{eq:normalized_infinity_matrix}
\end{equation}
Since $S^{-1}\partial_\phi S=-iJ$ and $S(\phi)=e^{-i\phi J}$,
$A_\infty$ commutes with $S(\phi)$. Let $\sigma_s$ and $P_s$ denote the
continuously labelled eigenvalue and Riesz projector approaching the spin
sector $s\in\{0,\pm1\}$. Standard spectral perturbation gives, uniformly in
$\phi$,
\begin{equation}
  \frac{\sigma_s(k,\phi)}{k^2}\longrightarrow i\beta s,
  \qquad
  P_s(k,\phi)\longrightarrow P_s(\infty),
  \qquad s\in\{0,\pm1\},
  \label{eq:normalized_polar_spectrum}
\end{equation}
whenever the relevant eigenvalue or cluster of $A_\infty$ is separated.
For the disk kernel, $\widehat G(k)=O(k^{-3/2})$, and the three branches have
the more explicit asymptotics
\begin{equation}
  \sigma_0(k)=O(k^{-7/2}),\qquad
  \operatorname{Re}\sigma_s(k)=O(k^{-3/2}),\qquad
  \operatorname{Im}\sigma_s(k)=\beta s k^2+O(1),
  \quad s=\pm1.
  \label{eq:unnormalized_polar_spectrum}
\end{equation}
Thus the real parts remain bounded and vanish asymptotically, whereas the
oscillatory pair is unbounded only through its imaginary parts. For every
continuously labelled branch, this asymptotic behaviour is independent of
the approach angle $\phi$. In the extended complex plane, each unbounded
branch reaches the spectral point at infinity independently of the direction
in wave-number space. This does not imply coalescence of the two oscillatory
branches in the affine complex plane: their separation grows as $O(k^2)$ and
their limiting projectors remain distinct. Therefore the target projector
extends to $S^2$ precisely in the nonsingular, gapped regimes for which these
polar separations hold.

\subsection{Spin Generator and Pole Formula}

Using
\begin{equation}
  \partial_kP_{\mathcal I}
  =S(\partial_kP_{\mathcal I}^{(0)})S^{-1},
  \qquad
  \partial_\phi P_{\mathcal I}
  =S[-iJ,P_{\mathcal I}^{(0)}]S^{-1},
  \label{eq:projector_polar_derivatives}
\end{equation}
the pullback of \eqref{eq:projector_chern} to $(k,\phi)$ is
\begin{equation}
  C(P_{\mathcal I})=\frac{1}{2\pi i}
  \int_0^\infty dk\int_0^{2\pi}d\phi\,
  \operatorname{Tr}\!\left\{
  P_{\mathcal I}^{(0)}
  [\partial_kP_{\mathcal I}^{(0)},[-iJ,P_{\mathcal I}^{(0)}]]
  \right\}.
  \label{eq:polar_chern_reduction}
\end{equation}
No additional factor of $k$ appears because
\eqref{eq:polar_chern_reduction} is the pullback of a differential
two-form, not a scalar area density.

To reduce the integrand, differentiate
$(P_{\mathcal I}^{(0)})^2=P_{\mathcal I}^{(0)}$:
\begin{equation}
  P_{\mathcal I}^{(0)}\partial_kP_{\mathcal I}^{(0)}
  +(\partial_kP_{\mathcal I}^{(0)})P_{\mathcal I}^{(0)}
  =\partial_kP_{\mathcal I}^{(0)},
  \qquad
  P_{\mathcal I}^{(0)}(\partial_kP_{\mathcal I}^{(0)})
  P_{\mathcal I}^{(0)}=0.
\end{equation}
Expanding the nested commutator and using cyclicity of the trace yields
\begin{align}
  &\operatorname{Tr}\!\left\{
  P_{\mathcal I}^{(0)}
  [\partial_kP_{\mathcal I}^{(0)},[-iJ,P_{\mathcal I}^{(0)}]]
  \right\}\nonumber\\
  &\hspace{2em}=\operatorname{Tr}\!\left[
  (-iJ)\left(
  P_{\mathcal I}^{(0)}\partial_kP_{\mathcal I}^{(0)}
  +(\partial_kP_{\mathcal I}^{(0)})P_{\mathcal I}^{(0)}
  \right)\right]
  =-i\operatorname{Tr}(J\partial_kP_{\mathcal I}^{(0)}).
  \label{eq:nested_commutator_identity}
\end{align}
The integrand is independent of $\phi$, so
\eqref{eq:polar_chern_reduction} becomes the compactification-pole formula
\begin{equation}
  C(P_{\mathcal I})=
  -\left[
  \operatorname{Tr}(J P_{\mathcal I}^{(0)}(k))
  \right]_{k=0}^{k=\infty},
  \label{eq:pole_chern}
\end{equation}
up to the overall sign fixed by the orientation convention.

The generator $J$ has eigenvectors
\begin{equation}
  e_\rho=(1,0,0)^T,\qquad
  u_+=(0,1,i)^T,\qquad
  u_-=(0,1,-i)^T,
\end{equation}
with spin weights $0,+1,-1$, respectively.
Define the corresponding rank-one projectors by
\begin{equation}
  Q_0=e_\rho e_\rho^\dagger,
  \qquad
  Q_s=\frac{u_su_s^\dagger}{u_s^\dagger u_s},
  \quad s=\pm1 .
\end{equation}
Then
\begin{equation}
  \operatorname{Tr}(JQ_s)=s,
  \qquad s\in\{0,+1,-1\}.
  \label{eq:spin_projector_weight}
\end{equation}
If a rank-$r$ projector at either compactified pole is a sum of such mutually disjoint spin
projectors, $\operatorname{Tr}(JP)$ is the sum of the included spin
weights. Equation \eqref{eq:pole_chern} therefore applies without
change to both a single band and a band cluster.

\subsection{Pole Eigenvectors and Twist Condition}

The pole formula becomes concrete by identifying which circular
polarization sector is selected at $k=0$ and at $k=\infty$. Let
$G_0=\widehat G(0)$ and
\begin{equation}
  b_0=
  -\omega+\frac{\lambda\rho_0}{2}G_0\sin\alpha .
  \label{eq:b0_pole}
\end{equation}
At $k=0$,
\begin{equation}
  B(0)=
  \begin{pmatrix}
  0&0&0\\
  0&a_0&b_0\\
  0&-b_0&a_0
  \end{pmatrix},
  \qquad
  a_0=\frac{\lambda\rho_0}{2}G_0\cos\alpha .
\end{equation}
Therefore
\begin{equation}
  B(0)u_+=(a_0+ib_0)u_+,\qquad
  B(0)u_-=(a_0-ib_0)u_- .
\end{equation}
For the branch selected by the lower-imaginary ordering, the $k=0$ pole spin
label is therefore
\begin{equation}
  s_0=-\operatorname{sgn}(b_0),
  \qquad
  k=0:\quad u_{s_0},
  \label{eq:zero_pole_sector}
\end{equation}
provided $b_0\ne0$.

At large $k$, the disk kernel satisfies $\widehat G(k)\to0$ and
\begin{equation}
  b(k)=\frac{v^2}{4D_0}k^2+O(1),
  \qquad
  D_0=2\omega-2\lambda\rho_0G_0\sin\alpha .
\end{equation}
Writing
\begin{equation}
  \beta=\frac{v^2}{4D_0},
\end{equation}
one obtains
\begin{equation}
  \frac{B(k)}{k^2}\longrightarrow
  A_\infty=
  \begin{pmatrix}
  0&0&0\\
  0&0&\beta\\
  0&-\beta&0
  \end{pmatrix},
\end{equation}
and hence
\begin{equation}
  A_\infty u_+=i\beta u_+,\qquad
  A_\infty u_-=-i\beta u_- .
\end{equation}
The same lower-imaginary ordering at infinity gives
\begin{equation}
  s_\infty=-\operatorname{sgn}(\beta),
  \qquad
  k\to\infty:\quad u_{s_\infty},
  \label{eq:infinity_pole_sector}
\end{equation}
provided $D_0\ne0$.

The target eigenvector therefore twists between opposite circular
polarizations exactly when the pole sectors in
\eqref{eq:zero_pole_sector} and \eqref{eq:infinity_pole_sector}
are different:
\begin{equation}
  \operatorname{sgn}(b_0)\ne\operatorname{sgn}(\beta)
  \quad\Longleftrightarrow\quad
  b_0\beta<0 .
  \label{eq:twist_condition_beta}
\end{equation}
Equivalently,
\begin{equation}
  D_0
  \left(
  -\omega+\frac{\lambda\rho_0}{2}G_0\sin\alpha
  \right)<0 .
  \label{eq:twist_condition_general}
\end{equation}
For the simulations in the main text, $\omega=0$. Then
\begin{equation}
  D_0b_0=-(\lambda\rho_0G_0)^2\sin^2\alpha ,
\end{equation}
so the pole sectors are opposite throughout $0<\alpha<\pi$.
They give a twist $u_-\to u_+$ for a globally isolated lower-imaginary
branch, but do not establish its finite-$k$ isolation. The phase-lag endpoints
$\sin\alpha=0$ are singular before this classification can be applied:
$D_0=0$, so $\beta=v^2/(4D_0)$ and the $D_0^{-1}$ entries of the closed
dispersion matrix are undefined. Consequently, at $\alpha=0$ and
$\alpha=\pi$ there is no finite closed matrix, no well-defined spectrum or
spectral projector, and hence no Chern number. These phase-lag endpoints must not be
interpreted as topologically trivial phases; the revised numerical scans
record them as unavailable rather than plotting them at zero.

Combining this pole-sector change with \eqref{eq:pole_chern},
the rank-one lower-imaginary-band projector obeys
\begin{equation}
  C=-\left(s_\infty-s_0\right),
  \qquad
  |C|=|(+1)-(-1)|=2,
  \label{eq:lower_imaginary_band_chern}
\end{equation}
where the displayed sign changes if the orientation of $S^2$ is reversed.
The upper-imaginary companion branch has the opposite pole-to-pole connection and hence
the opposite Chern number. For a two-band cluster complementary to a
rank-one projector,
\begin{equation}
  P_{\rm pair}=I_3-P_{\rm single},
  \qquad
  C(P_{\rm pair})=-C(P_{\rm single}),
  \label{eq:complementary_cluster_chern}
\end{equation}
because $\operatorname{Tr}J=0$. This explains why the single-band and
mixed-band numerical platforms carry the same magnitudes with
complementary signs. Thus $C=\pm2$ originates from the rotational
covariance of the dispersion matrix and the change of polar spin
weight, while the phase lag $\alpha$ controls whether that twist is
present through $D_0b_0$. All statements require $D_0\ne0$, the relevant
polar sectors to be nondegenerate, and the target projector to remain
isolated from its complement for every finite $k$. A straight bulk line
gap is an additional condition for the bulk--boundary identification below.

\section{UV-Regularized Toeplitz Strip and Boundary Spectral Flow}

\subsection{Normal-Momentum Regularization and Fourier Blocks}

Fix the tangential wave number $k_y$.  The continuum symbol
$M(k_x,k_y)$ is not periodic in $k_x$, so a finite Toeplitz matrix first
requires an explicit ultraviolet regularization.  Choose a normal-momentum
cutoff $k_c>0$, define the auxiliary lattice spacing
\begin{equation}
  a=\frac{\pi}{k_c},
  \qquad
  \Delta k_x=\frac{2k_c}{N_k},
  \qquad
  k_{x,j}=-k_c+j\Delta k_x,\qquad j=0,\ldots,N_k-1,
  \label{eq:strip_uv_grid}
\end{equation}
and periodically continue the sampled symbol from $[-k_c,k_c)$.  The
resulting Fourier blocks are
\begin{align}
  T_R(k_y)
  &=\frac{1}{2k_c}\int_{-k_c}^{k_c}
    M(k_x,k_y)e^{ik_xaR}\,dk_x,\nonumber\\
  &\approx\frac{1}{N_k}\sum_{j=0}^{N_k-1}
    M(k_{x,j},k_y)e^{ik_{x,j}aR},
  \qquad R\in\mathbb Z .
  \label{eq:strip_fourier_blocks}
\end{align}
Thus
\begin{equation}
  M_{k_c}^{\rm per}(k_x,k_y)
  =\sum_{R\in\mathbb Z}T_R(k_y)e^{-ik_xaR}
  \label{eq:periodized_strip_symbol}
\end{equation}
is the Fourier series of the \emph{periodized regularization}
$M_{k_c}^{\rm per}$, not an exact periodic identity for the original
continuum symbol.  This distinction is essential: $a$ is a numerical UV
scale and is unrelated to the particle interaction distance $d_0$.

The half-space block Toeplitz operator acts on
$\ell^2(\mathbb N_0)\otimes\mathbb C^3$ as
\begin{equation}
  (\mathcal T_{k_c}(k_y)\Psi)_n
  =\sum_{n'=0}^{\infty}T_{n-n'}(k_y)\Psi_{n'} .
  \label{eq:half_space_toeplitz}
\end{equation}
For the numerical strip, let $n,n'=0,\ldots,N_x-1$ label normal cells and
$\mu,\nu=1,2,3$ label the density--polarization components.  With a
Fourier-block cutoff $R_{\max}\leq N_x-1$, the hard-wall strip matrix is
\begin{equation}
  [\mathcal L_{\rm strip}(k_y)]_{(n,\mu),(n',\nu)}
  =
  \begin{cases}
    [T_{n-n'}(k_y)]_{\mu\nu},&|n-n'|\leq R_{\max},\\
    0,&|n-n'|>R_{\max}.
  \end{cases}
  \label{eq:strip_matrix_appendix}
\end{equation}
The open normal direction is therefore represented by truncating the
Toeplitz convolution, while translation invariance along $y$ leaves $k_y$
as a parameter.  Quantitative claims require convergence under increasing
$k_c,N_k,N_x,R_{\max}$; otherwise Fourier aliasing, the artificial UV
periodization, and hybridization between the two strip edges can move the
computed crossings.

\subsection{Left and Right Eigenvectors and Boundary Localization}

For each $k_y$, solve the non-Hermitian right and left eigenproblems
\begin{align}
  \mathcal L_{\rm strip}(k_y)\Psi_j^R(k_y)
  &=\sigma_j(k_y)\Psi_j^R(k_y),\nonumber\\
  \mathcal L_{\rm strip}(k_y)^\dagger\Psi_j^L(k_y)
  &=\sigma_j(k_y)^*\Psi_j^L(k_y),
  \label{eq:strip_left_right_eigenvectors}
\end{align}
and, for a simple spectrum, choose the biorthogonal normalization
\begin{equation}
  \langle\Psi_j^L(k_y),\Psi_l^R(k_y)\rangle=\delta_{jl}.
  \label{eq:strip_biorthogonal_normalization}
\end{equation}
Write the right eigenvector as
\begin{equation}
  \Psi_j^R=(\Psi_{j,0}^R,\ldots,\Psi_{j,N_x-1}^R),
  \qquad \Psi_{j,n}^R\in\mathbb C^3.
\end{equation}
Because $\Psi_j^R$ is the spatial profile multiplying the linear time
factor, the localization classifier uses its scale-invariant Euclidean
cell weight
\begin{equation}
  d_j^R(n;k_y)=\|\Psi_{j,n}^R(k_y)\|_2^2
  =\sum_{\mu=1}^3|\Psi_{j,n,\mu}^R(k_y)|^2.
\end{equation}
For an integer edge-window width $1\leq w<N_x/2$, define
\begin{equation}
  W_L^{(j)}=
  \frac{\sum_{n=0}^{w-1}d_j^R(n;k_y)}
       {\sum_{n=0}^{N_x-1}d_j^R(n;k_y)},
  \qquad
  W_R^{(j)}=
  \frac{\sum_{n=N_x-w}^{N_x-1}d_j^R(n;k_y)}
       {\sum_{n=0}^{N_x-1}d_j^R(n;k_y)}.
  \label{eq:strip_edge_weights}
\end{equation}
For a threshold $0<\eta<1$, a mode is labelled left-localized when
$W_L^{(j)}\geq\eta$ and $W_L^{(j)}>W_R^{(j)}$, and right-localized with
$L\leftrightarrow R$.  The threshold is a numerical classifier rather
than a topological invariant and must be varied in a robustness test.
Near an exceptional point or unresolved degeneracy, individual
biorthogonal eigenvectors cease to be stable labels. A subspace treatment
would then be required for invariant continuation \cite{Kunst2018}; the
present individual-branch implementation flags such samples and excludes
affected crossings rather than claiming to perform that continuation.

\subsection{Discrete Branch Tracking}

Let $k_y^{(s)}$, $s=0,\ldots,N_y-1$, be an increasing tangential-momentum
grid.  Eigenvalue ordering at a single grid point does not define a branch.
For adjacent samples, the implementation forms
\begin{equation}
  O_{jl}^{(s)}=
  \left|\left\langle
  \Psi_j^L(k_y^{(s)}),\Psi_l^R(k_y^{(s+1)})
  \right\rangle\right|,
  \qquad
  D_{jl}^{(s)}=
  \left|\sigma_j(k_y^{(s)})-\sigma_l(k_y^{(s+1)})\right|.
  \label{eq:branch_overlap_distance}
\end{equation}
After normalizing $O^{(s)}$ and $D^{(s)}$ by their respective maxima, the
assignment cost is
\begin{equation}
  \mathcal C_{jl}^{(s)}=
  (1-\chi)(1-\widetilde O_{jl}^{(s)})
  +\chi\widetilde D_{jl}^{(s)},
  \qquad 0\leq\chi\leq1.
  \label{eq:branch_assignment_cost}
\end{equation}
The permutation minimizing $\sum_j\mathcal C_{j,\pi(j)}^{(s)}$ is found by
the Hungarian assignment algorithm.  The reproduced strip plot uses
$\chi=0.25$.  This construction makes the branch labels depend mainly on
biorthogonal continuation while using eigenvalue proximity to resolve
finite-grid ambiguity. The code flags relative eigenvalue separations
at most $10^{-7}$, normalized left--right self-overlaps at most $10^{-8}$,
and assignment margins at most $10^{-6}$. The latter is the difference
between the best alternative row cost and the assigned cost; it is a
conservative individual-branch ambiguity flag, not a subspace-gap test.
Flagged crossings are recorded separately as unresolved.

\subsection{Line Gap and Oriented Crossings}

An oriented straight line in the complex spectral plane is specified by
\begin{equation}
  \ell=\{z\in\mathbb C:g(z)=0\},
  \qquad
  g(z)=\operatorname{Re}[e^{-i\theta}(z-z_0)].
  \label{eq:oriented_line_gap}
\end{equation}
It is a bulk line gap only when
\begin{equation}
  \operatorname{Spec}M(\mathbf k)\cap\ell=\varnothing
  \quad\text{for every finite }\mathbf k\in\mathbb R^2.
  \label{eq:bulk_line_gap_condition}
\end{equation}
The same separation must also persist in the high-$|\mathbf k|$ limiting
sector. The eigenvalue branches possess direction-independent asymptotic
values in the extended spectral plane, while the vector bundle over the
compactified wave-number space is extended through the uniform polar limits
of the Riesz projectors.
The sign of $g$ fixes the orientation convention.  For a tracked,
edge-localized branch, a transverse crossing at $k_y=k_*$ obeys
\begin{equation}
  g(\sigma_j(k_*))=0,
  \qquad
  \partial_{k_y}g(\sigma_j(k_y))|_{k_*}\neq0,
\end{equation}
and contributes
\begin{equation}
  s_{j,k_*}=\operatorname{sgn}
  \left[\partial_{k_y}g(\sigma_j(k_y))\right]_{k_*}.
  \label{eq:continuous_crossing_sign}
\end{equation}
On the discrete grid, a crossing is recorded when
$g_sg_{s+1}<0$, where $g_s=g(\sigma_j(k_y^{(s)}))$.  Linear interpolation
gives
\begin{equation}
  k_*\approx k_y^{(s)}+
  (k_y^{(s+1)}-k_y^{(s)})
  \frac{|g_s|}{|g_s|+|g_{s+1}|},
  \qquad
  s_{j,k_*}=\operatorname{sgn}(g_{s+1}-g_s).
  \label{eq:discrete_crossing_rule}
\end{equation}
The same edge label must hold at both bracketing samples; their average
weights are recorded only as diagnostics. A crossing exactly on a grid
vertex is bracketed once by its nearest nonzero neighbors, with the label
also required at intervening samples; tangencies and interval-end contacts
are not counted as transverse crossings. Crossings affected by a flagged
sample or ambiguous assignment are excluded and reported separately.
The resolved left and right counts are then
\begin{equation}
  \mathrm{SF}_{E,\ell}
  =\sum_{(j,k_*)\in\mathcal X_E}s_{j,k_*},
  \qquad E\in\{L,R\},
  \label{eq:discrete_edge_spectral_flow}
\end{equation}
where $\mathcal X_E$ is the set of transverse crossings classified at edge
$E$.  Horizontal lines $\operatorname{Im}\sigma=c$, as used in the strip
figure, correspond to $\theta=\pi/2$ and $z_0=ic$.

For the continuum problem, the tangential parameter is
$k_y\in\mathbb R\cup\{\infty\}\simeq S^1$.  A finite interval
$[-k_{y,c},k_{y,c}]$ approximates this compactification only if the two
ends have reached the same high-$|k_y|$ limiting sector and no crossing is
lost near either cutoff.  Otherwise
\eqref{eq:discrete_edge_spectral_flow} is only a cutoff-dependent crossing
count, not an integer spectral-flow invariant.

\subsection{Scope of the Bulk--Boundary Identification}

For a Hermitian Chern family, the Toeplitz/Fredholm index identifies the
oriented edge spectral flow with the Chern-number difference across an
interface \cite{Kellendonk2002,GrafPorta2013,ProdanSchulzBaldes2016}.
A line-gapped non-Hermitian family admits a gap-preserving flattening to a
Hermitian representative \cite{Kawabata2019}.  Nevertheless, conventional
Bloch bulk--boundary correspondence can fail in non-Hermitian systems with
skin accumulation, and left, right, and biorthogonal localization need not
coincide \cite{Kunst2018,YaoWang2018}.  Accordingly, the strip calculation
here uses
\begin{equation}
  \mathrm{SF}_{\rm interface,\ell}
  =C(P_{\rm right})-C(P_{\rm left})
  \label{eq:bulk_boundary}
\end{equation}
as a consistency relation only when: (i) the same bulk line gap remains
open; (ii) the compactified bulk projector is continuous; (iii) the
half-space family is Fredholm; (iv) the UV-regularized strip converges
without extensive bulk skin accumulation; and (v) the $k_y$ cutoff closes
onto the same limiting sector.  For a hard wall modelled by a trivial
exterior, $C(P_{\rm vac})=0$.  Opposite boundary orientations then give
\begin{equation}
  \mathrm{SF}_{L,\ell}=C(P),
  \qquad
  \mathrm{SF}_{R,\ell}=-C(P)
  \label{eq:opposite_edge_flows}
\end{equation}
up to one global orientation sign.  A finite-strip agreement with
\eqref{eq:opposite_edge_flows} is numerical evidence for bulk--boundary
correspondence; it is not, by itself, a proof that all of the above
operator-theoretic hypotheses hold.

\subsection{From Edge Spectrum to Boundary Motion}

An edge wave packet built narrowly around a real $k_y=\bar k_y$ contains
modes
\begin{equation}
  U(x,y,t;k_y)=\Psi^R(x;k_y)e^{\sigma(k_y)t+ik_y y},
  \qquad
  \sigma(k_y)=\gamma(k_y)+i\Omega(k_y).
\end{equation}
With the convention $e^{\sigma t+ik_yy}$, its oscillatory phase is
$k_yy+\Omega(k_y)t$.  If $\gamma(k_y)$ and the eigenvector profile vary
slowly across the packet bandwidth, stationary phase gives
\begin{equation}
  v_g=\frac{dy}{dt}=-\Omega'(\bar k_y).
  \label{eq:edge_group_velocity}
\end{equation}
This sign changes if the Fourier convention or boundary orientation is
reversed.  Nonzero spectral flow enforces a net oriented crossing count;
it does not require every edge branch to have the same local group velocity
and does not prove nonlinear particle-flow stability.  The observed
particle flow is therefore compared with the linear edge branches only
after nonlinear saturation and persistent boundary localization have been
checked independently.

\section{Numerical Workflow}

The reproducible workflow is:
\begin{enumerate}[label=\arabic*.]
  \item Fix $M(k_x,k_y)$ and the regularization parameters
  $(k_c,N_k,R_{\max},N_x)$ using \eqref{eq:strip_uv_grid}--
  \eqref{eq:strip_matrix_appendix}.
  \item At every $k_y^{(s)}$, construct $\mathcal L_{\rm strip}$ and compute
  paired left and right eigenvectors satisfying
  \eqref{eq:strip_biorthogonal_normalization}.
  \item Track simple branches by minimizing
  \eqref{eq:branch_assignment_cost}; flag unresolved eigenvectors and
  ambiguous assignments, without substituting an unimplemented subspace rule.
  \item Compute \eqref{eq:strip_edge_weights} and classify crossings only
  when the edge label persists on both sides of the crossing.
  \item Count resolved oriented crossings of the stated empirical reference
  lines using \eqref{eq:discrete_crossing_rule}, and separately test whether
  each reference intersects the bulk spectrum. The line-gap condition
  \eqref{eq:bulk_line_gap_condition} is required only for an invariant
  bulk--boundary interpretation, not for reporting these empirical counts.
  \item Repeat after varying $k_c,N_k,R_{\max},N_x,w,\eta,\Delta k_y$, and
  $k_{y,c}$.  Report an integer spectral flow only when its value and the
  crossing locations are stable under these changes.
  \item Compare the converged edge counts with the bulk Chern number from
  the biorthogonal Fukui projector algorithm.
\end{enumerate}
The circular-particle-matched strip calculation uses
$N=2000$, $L=7$, $R=L/2=3.5$, $v=3$, $\omega=0$, $K=20.75$, and $d_0=1$,
with $\rho_0=N/(\pi R^2)$ and $\lambda=K/(\rho_0\pi d_0^2)$.
The sampled phase lags are $\alpha/\pi=0.2,0.4,0.5,0.6,0.8$; neither
singular endpoint is evaluated. The displayed numerical grid is
\begin{equation}
  (k_c,N_k,R_{\max},N_x,w,\eta,k_{y,c},N_y,\chi)
  =(40,384,35,36,6,0.45,50,201,0.25).
  \label{eq:strip_reproduction_parameters}
\end{equation}
The levels $\operatorname{Im}\sigma=\pm10$ are empirically chosen observation
references. For $\alpha=\pi/2$, the matched bulk spectrum is
$0,\pm i\sqrt{b(k)^2+v^2k^2/2}$ with minimum nonzero frequency
$K/2=10.375$, so both levels are bulk line gaps; for the other four phase
lags they intersect the bulk spectrum and remain only observation references.
Finite-strip counts, including the threshold case, do not by themselves
certify every convergence assumption of the index relation. The $N_y=101$
run is a momentum-step comparison, not a joint UV or infinite-width convergence
claim. On both grids, all five phase lags give
$(\mathrm{SF}_{L,\ell},\mathrm{SF}_{R,\ell})=(2,-2)$ at both references under
the stated orientation and persistent-edge rule. The machine-readable output
records parameters, resolved and excluded crossings, and bulk intersections
for every phase lag.

\begin{figure}[htbp]
 \centering
 \includegraphics[width=\textwidth]{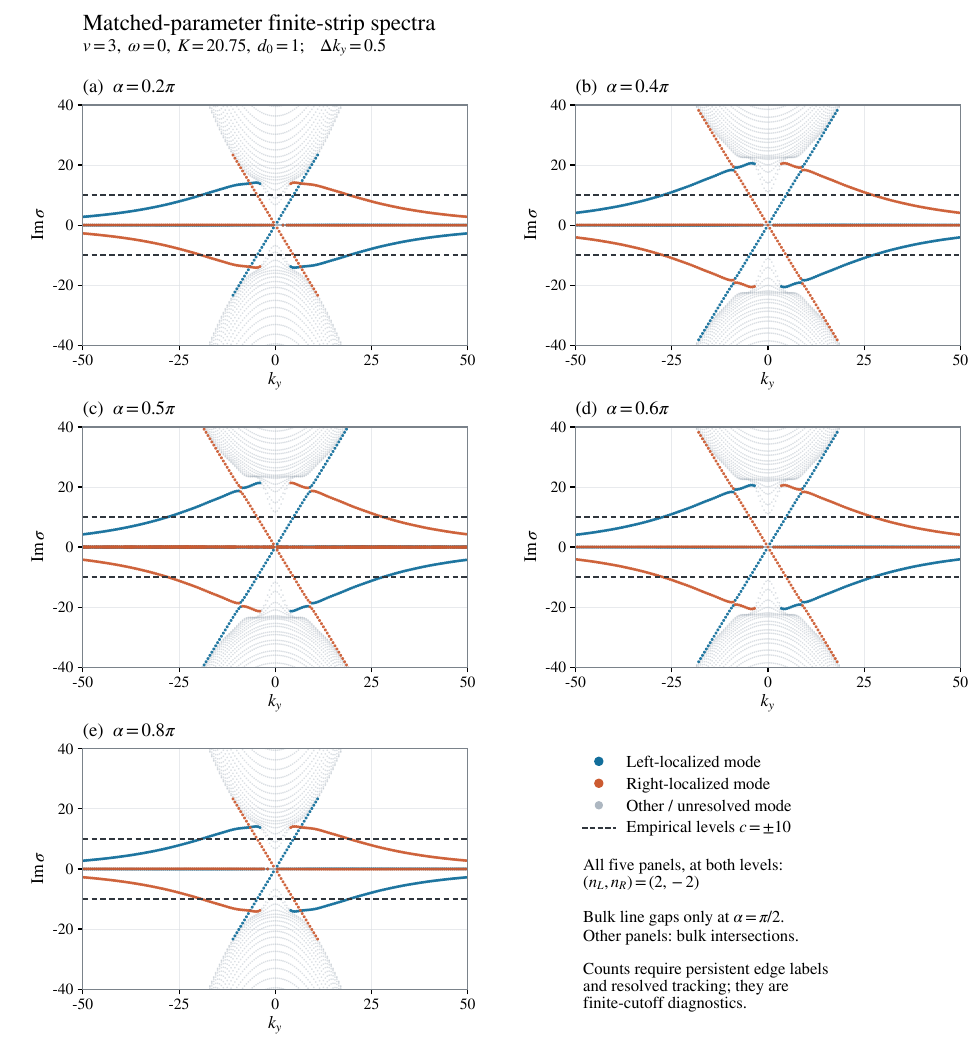}
 \caption{Finite-strip spectra at the circular-particle-matched parameters.
 Blue and orange samples are left- and right-localized modes; gray samples
 are other or unresolved modes. The two unchanged empirical reference
 levels are $\operatorname{Im}\sigma=\pm10$. Displayed signed counts require
 persistent localization on both sides of a crossing and resolved tracking;
 they are finite-cutoff diagnostics, not certified line-gap invariants.
 All panels share the same axes and numerical grid.}
 \label{fig:strip_circle_matched_sweep}
\end{figure}

\clearpage
\section{Summary}

Starting from the microscopic frustrated alignment rule, the kinetic
equation is reduced to a closed density--polarization model by Fourier
mode truncation and adiabatic elimination of the second angular harmonic.
Linearization around the homogeneous state gives a non-Hermitian
dispersion matrix whose real parts determine growth rates and whose
imaginary parts determine linear frequencies. Isolated spectral
projectors of this matrix define complex vector bundles over wavenumber
space; their first Chern numbers are computed either through the
projector formula or through the Fukui determinant-link discretization.
This closed matrix and its topological invariants require $D_0\ne0$; for
$\omega=0$, the phase-lag endpoints $\alpha=0$ and $\alpha=\pi$ are singular and
carry no spectrum or Chern number within the closure.
Rotational covariance reduces the relevant Chern number to a difference
of polar spin weights, explaining the robust values $C=\pm2$. Finally,
Toeplitz strip matrices convert the bulk symbol into an open-boundary
linear problem. Their empirical edge branches are compared with particle
transport; identifying a crossing count with a bulk Chern number additionally
requires the common line-gap and convergence conditions stated above.